\documentclass[11pt,english]{article}
\pdfoutput=1
\usepackage{jheppub}
\usepackage{lscape}
\usepackage{amsmath}
\usepackage{slashed}
\usepackage{mathtools}
\usepackage{tikz}
\usetikzlibrary{shapes,arrows,shadows}
\usepackage{amsmath,bm,times}
\usepackage{hyperref}
\usepackage{cleveref}
\usepackage{tcolorbox}

\def\-{\hphantom{-}}

\def\s2{\frac{1}{\sqrt2}}

\def\be{\begin{equation}}
\def\ee{\end{equation}}
\def\bea{\begin{align}}
\def\eea{\end{align}}
\def\beqa{\begin{eqnarray}}
\def\eeqa{\end{eqnarray}}

\def\tr{{\rm tr \,}}

\def\mg{m_{3/2}}
\def\mg2{m^2_{3/2}}

\def\Dsl{\,\raise.15ex\hbox{/}\mkern-13.5mu D} %this one can be subscripted

\def\H{{\text{H}}}
\def\eV{\mathcal{V}_{\text{eff}}}

\newcommand{\eq}[1]{\begin{equation}
                     \begin{split} #1 \end{split}
                     \end{equation}}

\begin{document}

\title{Metastable vacua from torsion and machine learning}% Force line breaks with \\

\author[a]{Cesar Damian} 
\author[b]{and Oscar Loaiza-Brito}

\affiliation[a]{Departamento de Ingenier\'ia Mec\'anica, Universidad de Guanajuato, Carretera Salamanca-Valle de Santiago Km 3.5+1.8 Comunidad de Palo Blanco, Salamanca, Mexico.}
\affiliation[b]{Departamento de F\'isica, Universidad de Guanajuato, Loma del Bosque No. 103 Col. Lomas del Campestre C.P 37150 Leon, Guanajuato, Mexico.}

%\emailAdd{\ldots}

\date{\today}% It is always \today, today,
             %  but any date may be explicitly specified

\abstract{
By implementing an error function on a Machine Learning algorithm we look for minimal conditions to construct stable Anti de Sitter and de Sitter vacua  from dimensional  type IIB String theory compactifcation on  K\"ahler manifolds with torsion. This allows to have contributions to the scalar potential from the five-form flux and from D-branes wrapping torsional cycles, interpreted as non-BPS states. The former implies the possibility to construct stable AdS vacua while the later constitutes a mechanism to uplift AdS to dS vacua. Particularly we consider $\hat{D5}$ non-BPS states to uplift the stable AdS vacua to an (apparently) stable dS minimum. Both results $-$the generation of an AdS vacuum and the corresponding uplifting  to a dS one$-$ are restricted to certain type of configurations, specifically with the number of $O3$ orientifolds bounded from below by  the number of $D3$-branes and  fluxes.
Under these conditions we report over 170 dS  (classical) stable vacua. In all of them, the uplifted effective potential becomes very flat indicating the presence of possible source of instabilities. We comment about their relation with the Swampland Conjectures.\\
}
\arxivnumber{}


\maketitle

\section{Introduction}
The Swampland Program has received a lot of attention over the last  few years. Its importance relies on the establishment of some criteria to  separate effective quantum field theories $-$considered as consistent with Quantum Gravity, a.k.a. String Theory$-$ from those which are not.  The  program focuses in  different proposals commonly referred as Conjectures which appear to rule out some of the string model engineering constructions so far presented in the literature.  Some of those conjectures are involved in our work, such as  the instability of non-SUSY Anti de Sitter (AdS) vacua, the AdS scale separation and the Refined de Sitter conjecture, which in turn seem to be interconnected \cite{Palti:2019pca, vanBeest:2021lhn, Grana:2021zvf}).  \\

%The Swampland criteria appears to rule out a plethora of  string model engineering constructions so far proposed in the literature.  Several proposals of the swampland program put some usual compactification scenarios in string theory, that try to address some deep questions about the early Universe, out of the landscape.
%all the well-known compactification scenarios in string theory in tension with observational data as well as with phenomenological constraints, 
%In this work we are interesting in  the construction of  (meta)stable de Sitter (dS) vacua from a string compactification  at the perturbative level and how such vacua could be unstable or genercially have some properties which put them into the Swampland. Particularly we shall concentrate on compactifications on manifold which allow to have the minimal contributions to the effective scalar potential which allow to have a dS vacuum. \\

The refined dS conjecture establishes that the minima of the scalar potential coming from the dimensional reduction of the low energy theory in string theory have to be AdS otherwise they are tachyonic or not consistent to Quantum Gravity, at least in the asymptotic regions of moduli space \cite{Ooguri:2006in, Garg:2018reu, Ooguri:2018wrx,Lust:2019zwm, Apers:2022zjx}. Even more restrictive, the AdS conjecture establishes that the scale of the lightest moduli is not parametrically separated from the AdS scale, and thus any attempt to uplift an AdS to a dS vacuum shall result in their destabilization\footnote{See \cite{Bena:2018fqc} for the case in the deformed conifold.}. \\

 Recently it has been argued that the use of non-BPS states, classified by K-theory, shall be an interesting corner to evade these restrictions \cite{Blumenhagen:2019kqm,Damian:2019bkb}, unless the total K-theory charge must cancel as pointed out in \cite{Uranga:2000xp} and related to the cobordism conjecture in \cite{Blumenhagen:2021nmi}. 
%The non-BPS states of interest in this project which are classified by K-theory are invisible to the cohomology and can be constructed by a bound state of brane - anti brane states in presence of an orientifold projection \cite{Sen:1999mg,Sen:1998ii,Sen:1998rg,Loaiza-Brito:2001yer,Witten:2000cn}. Some of the consequences of having such states has been recently studied, concluding that the presence of discrete K-theory charge violates the refined dS conjecture for a non-zero total K-theory charge over the compact internal manifold in compactificacion in the type IIA string theory .\\
%
%Besides the non-BPS which carries tachyonic states related to strings stretching between the mixed sector of non-BPS and BPS states which is argued to be stable against decaying to a vacuum configuration of the theory. To understand this stability, it is considered the non-BPS $\widehat{D}7$-brane in Type I theory which carries K-theory $\mathbb{Z}_2$ charge, which is not in the same K-theory class as the vacuum state \cite{Loaiza-Brito:2001yer}, thus the decay shall proceed to a state within the same K-theory class. For instance, in the compactified limit the final states reaches a toron gauge configuration which vanishing Chern classes and non-trivial $\mathbb{Z}_2$ charge. \\
The use of non-BPS states, typically constructed from a pair of stable branes and anti-branes in the presence of an orientifold plane, emulates the role played by non-perturbative contributions in KKLT scenarios 
%In the type IIB compactifications, it is notoriously hard to obtain a dS vacua with all moduli stabilized. For instance in the KKLT scenario, the stabilization of the K\"ahler moduli require the presence of a minimal set of ingredients which includes non-perturbative contributions which breaks 
by breaking the non-scale structure of the $\mathcal{N}=1$ superpotential and providing a nice mechanism to stabilize all the moduli. 
%In analogy, the presence of non-BPS $\widehat{D}p$ branes breaks the non-scale structure providing a mechanism to stabilize the moduli without requiring non-perturbative information of the theory. 
%However, even by including the F$_3$ as well as H$_3$ fluxes treading appropriate cycles of the compact space it is not possible to stabilize all the moduli. Thus, it is necessary to carry out a systematic study of the ingredients required to stabilize all the moduli by considering a more complete set of elements that can contribute to the scalar potential, \emph{i.e.}, F$_1$, $F_5$, curvature contributions, among other sources. 
However, their inclusion is not  suffice to guarantee the presence of apparently stable dS vacua but  contributions to the effective scalar potential coming from the RR 5-form are necessary.\\ 

We are interested in two main aspects. First, in constructing a (meta)-stable dS vacuum by identifying the minimal set of ingredients the effective scalar potential must posses in the spirit of \cite{Hertzberg:2007wc, Shiu:2011zt} and also to find possible compactification scenarios where such conditions might be present. Second, in case we can construct a classical stable dS vacuum we want to look for possible sources of instabilities which in turn can be taken as evidence (or not) of the realization of the above referred Swampland Conjectures .\\

In this work we consider a compactification on a K\"ahler manifold admiting torsion, upon which there is a contribution of the torsional part of  $F_5$ to the scalar potential allowing to find AdS vacua (but not dS).  For that to happen it is necessary that the number of orientifold fixed points  be greater than the number of $D3$-branes such that their contribution to the tadpole be negative, i.e., $\mathcal{N}_3<0$. Under this context it is then possible to wrap $D5$-branes on torsional 2-cycles which we claim are precisely $\hat{D5}$ non-BPS states and that contribute with a positive amount of energy such that  uplift the AdS vacuum to a dS one.\\

As in the case of the AdS vacua, the realization of dS minima require some extra conditions, namely that there are fluxes in RR and NS-NS sectors supported in more than two 3-cycles and that the number of orientifold 3-planes has a lower bound given by
\begin{equation}
\mathcal{N}_{O3}>4\frac{ (A_{H_3} A_{F_3})^{1/2}}{A_3}+2\mathcal{N}_{\text{flux}},\nonumber
\end{equation}
where $A_{H_3}, A_{F_3}$ and $A_3$ are the contributions $-$upon dimensional reduction$-$ of 3-form fluxes and 3-dimensional sources as $D3$-branes and $O3^-$-planes, while $\mathcal{N}_{\text{flux}}$ is the usual flux number entering into the $D3$-brane charge tadpole contribution.\\

These conditions were inferred after implementing a Machine Learning (ML) algorithm specifically designed to look for dS vacua. The use of ML algorithms and tools  has been proved to be prolific (and a more systematic way)  to explore the vacua in string theory compactifications (see for instance \cite{He:2018jtw, Ashmore:2019wzb, Parr:2019bta, Bao:2020sqg, Halverson:2020opj, Gal:2020dyc,  Erbin:2020tks, CaboBizet:2020cse, Cole:2021nnt, He:2022cpz}).  For that we implemented a hybrid algorithm to explore the minima of a  scalar potential of the form\footnote{Other sources are considered in the Appendix \ref{ML} such as fluxes, branes and  negative curvature. More exotic fluxes, as non-geometric has been considered in the literature (see \cite{Plauschinn:2018wbo} and references therein). }
\begin{equation}
\eV = \eV( {H}_3, {F}_3, {F}_5, \widehat{D5})\nonumber
\end{equation}
subject to the constraints of 1) having a positive value at the minimum, 2) zero value of its derivative with respect to each of the moduli, 3) positive definiteness of the mass matrix, and 4) positiveness of the contribution of the $ \widehat{\text{D}5}$-brane. In the context of ML, these restrictions can be implemented through an objective function  written as
\begin{equation}
\text{Error} = \sum_{i=1}^N \alpha_i \text{error}_i\nonumber
\end{equation} 
where each of the error$_i$ contributions takes into account every single  restriction above mentioned with the $\alpha$ parameter a real value giving a weight to each error contribution.  In the present work we employ the hybrid algorithm including the Simulated Annealing (SA) as well as the Gradient Descent (GD). The SA algorithm is a heuristic method for solving optimization problems which, inspired by the annealing procedure of metal working,  is able to look for an approximate solution to the optimization problem. The GD algorithm on the other hand is a second order iterative optimization algorithm designed to find local minima provided that the first derivative is known. Thus, at a first step, the SA algorithm shall look for interesting points in the error function whereas the GD shall improve the solution guaranteeing the zero value of the first derivative of the scalar potential. We describe in detail these algorithms in Appendix \ref{ML}.\\

Our work is then organized as follows: In section 2 we present the most usual conditions for a type IIB compactification and specify the notation we use along the paper. In section 3  we show that it is possible to construct AdS vacua by compactification of type IIB string Theory on a K\"ahler manifold with torsion, such that the RR five-form has a torsional contribution to the effective scalar. For that we implement a ML algorithm  through the presence of an error function which allows to easily find a large number of stable and unstable vacua. In this case we report  389 different AdS vacua which existence relies upon the requirement that the number of $D3$-branes be less than one-half of the number of orientifold $O3^-$-planes.  However no dS vacua were found under these conditions. In section 4, once we take a compactification on a manifold with torsion,  we also consider $D5$-branes wrapping torsional 2-cycles  while fulfilling the aforementioned conditions on fluxes and the orientifold bound.  Extra assumptions were taken, such as the non existence of torsional components of all 3-form fluxes.  For this case, we report over 170 different dS stable vacua. In section 5 we discuss the conditions upon which the AdS vacua can be lifted to dS ones and comment about the implications with respect to the Swampland Conjectures. In section 6 we present our conclusions, while in the Appendix we describe some useful technical information in relation with the Machine Learning algorithm to be implemented in our search, particularly about the incorporation of the above mentioned two algorithms: the Simulated Annealing and the Conjugate Gradient.\\

\section{Contribution to the scalar potential}
Let us review the standard dimensional reduction procedure to construct the effective scalar potential. Consider the type IIB superstring compactified on a manifold $\mathbb{X}_6$ in the presence of 3-form fluxes and 3-dimensional local sources. We are not including 7-branes or orientifold 7-planes. As usual, the action for the massless modes in the string frame is
\begin{equation}
S_{IIB}= S_{G}+S_\phi+S_{G_3}+S_{F_5}+S_{CS}+S_{\text{loc}},
\end{equation}
with
\begin{eqnarray}
S_{G}&=&\frac{1}{2\kappa^2_{10}}\int d^{10}x\sqrt{-G}~e^{-2\phi}R,\\
S_{\phi}&=&\frac{1}{2\kappa^2_{10}}\int d^{10}x\sqrt{-G}\left[e^{-2\phi}\left(4(\nabla\phi)^2\right)-\frac{|F_1|^2}{2}\right],\\
S_{G_3}&=-&\frac{1}{4\kappa^2_{10}}\int d^{10}x\sqrt{-G} \left(e^{-2\phi}|H_3|^2-|\hat{F}_3|^2\right),\\
S_{F_5}&=&-\frac{1}{8\kappa^2_{10}}\int d^{10}x\sqrt{-G}~|F_5|^2,\\
S_{CS}&=&-\frac{1}{4\kappa^2_{10}}\int C_4\wedge H_3\wedge F_3,\\
S_{\text{loc}}&=&S_{\text{DBI}}+S_3=T_3\mathcal{N}_3\int d^4x \sqrt{-g_4} e^{-2\phi} +\frac{1}{2}\mathcal{N}_3~\mu_3\int_{\Sigma_4} C_4,
\end{eqnarray}
where in terms of the string length $l_s$, 
\begin{equation}
\kappa^2_{10}=\frac{l_s^8}{4\pi},
\end{equation}
$T_3$ is the $D3$-brane tension, $\mathcal{N}_3= \mathcal{N}_{D3}-\frac{1}{2}\mathcal{N}_{O3}$ counts the number of $D3$-branes minus the number of orientifold planes $O3^-$ with $\mu_3=T_3=\frac{2\pi}{l_s^4}$.  We consider the DBI action at leading order in $\alpha'$ for $D3$-branes and $O3^-$-planes along the extended coordinates, where
 the RR fluxes are 
\begin{eqnarray}
\hat{F}_3&=&F_3-C_0\wedge H_3,\nonumber\\
F_5&=&dC_4-\frac{1}{2}C_2\wedge H_3-\frac{1}{2}B_2\wedge dC_2.
\end{eqnarray}
Thus, the action $S_{F_5}$ (before self-duality is imposed) can be written as
\begin{equation}
S_{F_5}=-\frac{5!}{8\kappa^2_{10}}\int F_5\wedge\ast F_5= \frac{15}{\kappa^2_{10}}\int \left[ C_4\wedge d\ast F_5 + (\frac{1}{2}C_2\wedge H_3+\frac{1}{2}B_2\wedge dC_2)\wedge \ast F_5\right].
\label{S5}
\end{equation}
Due to the action of the orientifold planes $O3^-$, the RR and NS-NS potentials $C_2$ and $B_2$ are projected out and the equations of motion from $\delta S/\delta C_4=0$ give us the tadpole condition for the 3-dimensional sources
\begin{equation}
\mathcal{N}_3+\frac{1}{l_s^4}\int F_3\wedge H_3=\mathcal{N}_3+\mathcal{N}_{\text{fluxes}}=0.
\end{equation}
Therefore, the contribution from $S_{F_5}+S_{CS}+S_3$ to effective the scalar potential $-$in a compactification on a CY manifold$-$ vanishes. As we shall see we are going to depart from a CY compactification into a more general setup such that $S_{F_5}$ does have a contribution.\\

In order to construct the effective scalar potential $\eV$, we specify the  ten-dimensional metric as
\begin{eqnarray}
ds_{10}^2&=&g_{\mu\nu}dx^\mu dx^\nu+h_{mn}dy^ndy^m,\nonumber\\
&=& e^{-2\Omega}e^{2A(y)}\tilde{g}_{\mu\nu}dx^\mu dx^\nu+ e^{-2A(y)}\tilde{h}_{mn}dy^mdy^n,
\label{metric}
\end{eqnarray}
where $e^{-2\Omega}$ is the conformal factor fixed as
\begin{equation}
e^{-2\Omega}= e^{-2\phi} \mathcal{V}_6
\end{equation}
to change into the Einstein frame, with $\mathcal{V}_6=\int d^6y\sqrt{h_6}$.
%We also consider the presence of  3-dimensional sources  for $F_5$ given by
%\begin{equation}
%S_3=\frac{1}{2}\mathcal{N}_3~\mu_3\int_{\Sigma_4} C_4,\\
%\end{equation}
%where $\mathcal{N}_3= \mathcal{N}_{D3}-\frac{1}{2}\mathcal{N}_{O3}$ counts the number of $D3$-branes minus the number of orientifold planes $O3^-$ with
%\begin{equation}
%\mu_3=T_3=\frac{2\pi}{l_s^4}.
%\end{equation}
%
%after imposing the tadpole condition, reduces to
%\begin{equation}
%S_{F_5+CS+3}= \frac{15}{2\kappa^2_{10}}\int \omega_5\wedge \ast F_5.
%\end{equation}
%where
%\begin{equation}
%\omega_5=C_2\wedge H_3+\frac{1}{2}B_2\wedge F_3.
%\end{equation}
%
%Our first step is to compute the effective contribution to the scalar potential of a Type IIB compactification on a CY manifold $\mathbb{X}_6$ threatened with fluxes and in the presence of $D3$-branes and $O3^-$-planes with $\mathcal{N}_3<0$ and $\mathcal{N}_{fluxes}>0$.\\
%
%In order to compute their contributions, we change into the Einstein frame without taking a stabilized dilaton. This is, we express the Einstein-Hilbert action as
%\begin{equation}
%S_g=\frac{1}{\kappa^2_{10}}\int d^4x \sqrt{-g_4} R(g),
%\end{equation}
%where  $g_{\mu\nu}$ in Eq.(\ref{metric}) is the metric in the Einstein frame  with the conformal factor given by
%\begin{equation}
%e^{-2\Omega}= e^{2\phi} \mathcal{V}_6
%\end{equation}
%and
%\begin{equation}
%\mathcal{V}_6=\int d^6y\sqrt{h_6}
%\end{equation}
Notice we are not taking into account warping effects on the internal metric. \\

In terms of the axionic moduli fields $\tau$ and $s$ are given by
\begin{equation}
\tau=e^{-\phi}(\mathcal{V}_6)^{2/3}, \qquad s=e^{-\phi},
\end{equation}
so, the contributions for the action terms $S_{G_3}$ and $S_{DBI}$ are given by
\begin{eqnarray}
%S_{f5}&=&\int d^4x \sqrt{-g_4} ~\frac{A_5}{\tau^4}=0 \qquad\text{for a CY manifold},\label{f5}\\
S_{G_3}&=&\int d^4x\sqrt{-g_4} ~\left(\frac{A_{F_3}}{s\tau^3}+\frac{A_{H_3}s}{\tau^3}\right),\label{g3}\\
%S_{H_3}&=&\int d^4x\sqrt{-g_4} ~\frac{A_{H_3}s}{\tau^3},\label{h3}\\
S_{\text{DBI}}&=&\int d^4x\sqrt{-g_4} ~\frac{A_3 \mathcal{N}_3} {\tau^3}\label{dbi},
\end{eqnarray}
where $A_{F_3}, A_{F_3}$ and $A_3$ are the corresponding contributions not depending on  $\tau$ and $s$ where 
\begin{equation}
S=C_0+is, \qquad \text{and} \qquad T= \int C_4 + i\tau.
\end{equation}

On the above we have assumed
 that complex structure moduli $z_i$ are  fixed through 3-form fluxes, by  $D_{z_i}\mathcal{W}=0$, where as usual
\begin{equation}
\mathcal{W}=\int (F_3-SH_3)\wedge\Omega(z_i),
\end{equation}
but $D_S\mathcal{W}\ne 0$. Therefore SUSY is broken at least by the axio-dilaton modulo $S$, and the fluxes we are turning on,  have not $(1,2)$-components. Together with the K\"ahler potential of the form
\begin{equation}
\mathcal{K}=-\log(-i(S-\bar{S}))-3\log(- i (T - \bar T )),
\end{equation}
the flux contribution to the scalar potential reduces to
\begin{equation}
\mathcal{V}_{\text{fluxes}}=e^{\mathcal{K}}|D_SW|^2K^{S\bar{S}}=\frac{\hat{f}^2+s^2h^2}{2s\tau^3}.
\end{equation}
with 
\begin{equation}
\hat{f}=\int\hat{F}_3\wedge \Omega, \qquad \text{and}\qquad h=\int H_3\wedge\Omega.
\end{equation}
Comparing with expression (\ref{g3}), 
\begin{equation}
A_{F_3}=\frac{|\hat{f}|^2}{2\kappa_{10}^2}, \qquad \text{and}\qquad A_{H_3}=\frac{|h|^2}{2\kappa_{10}^2}.
\end{equation}

As known, by exploring different values for $A_{F_3}$, $A_{H_3}$ and $A_3$ we find that no stable vacuum is obtained. More ingredients are required.

%-------------------------------------------------------------

\section{Stable non SUSY AdS vacua from torsion}

As suggested in the  literature (see  \cite{Hertzberg:2007wc, Shiu:2011zt} and \cite{Danielsson:2012et, Blumenhagen:2015xpa, Junghans:2016uvg, CaboBizet:2016qsa, Andriot:2018ept, Kallosh:2018nrk, Andriot:2018mav, Andriot:2019wrs, Andriot:2020wpp, Andriot:2021rdy}), it is possible to find stable vacua by turning on different contributions to the scalar potential.
Here we are interested in a non-vanishing contribution  from $S_{F_5}$ to $\eV$. For that,  we shall take into account the presence of torsion in the internal manifold $\mathbb{X}_6$ which, as we shall argue,  naturally comes into play in the presence of orientifold planes \cite{McAllister:2008hb, Cai:2014vua}. This implies that the K\"ahler 2-form $J_2$ is no longer closed, i.e., $dJ_2\ne 0$ pointing out the necessity to compactify on generalized CY manifolds. By using the ML algorithm described in the Appendix \ref{ML}, we find that AdS stable vacua are obtained under  some specific  conditions we shall describe in detail.

\subsection{Effective scalar potential from Torsion}

Let us start by writing the  action component $S_{F_5}$ in (\ref{S5}) as
\begin{equation}
S_{F_5}= \frac{15}{2\kappa^2_{10}}\int \omega_5\wedge \ast F_5.
\end{equation}
where
\begin{equation}
\omega_5=\frac{1}{2} C_2\wedge H_3+\frac{1}{2}B_2\wedge F_3.
\end{equation}
As said, in generic compactifications on $\mathbb{X}_6$ with orientifold planes $O3^-$, 2-forms are 
divided on odd or even  according to the orientifold action on them \cite{Grimm:2004uq}. Since  2-form RR and NS-NS potentials are odd under an $O3^-$ action,  and the fluxes  $F_3,H_3$ are even and it follows that
\begin{equation}
\omega_5\in \Omega^2_-( \mathbb{X}_6,\mathbb{Z})\wedge H^3_+( \mathbb{X}_6,\mathbb{Z}).
\end{equation}
Therefore, for a generic CY manifold, $\omega_5$ does not contribute to $\eV$.
%Since
%\begin{equation}
%d:\Omega^2_-\longrightarrow [0]\in \text{H}^3(X_6),
%\end{equation}
%then $S_{f5}=0$.
 Also notice that in  the presence of orientifold $O3^-$-planes, the RR potential $C_6$ is projected out and it is not possible to have stable BPS $D5$-branes. The effective 4-dimensional scalar potential only receives contributions from the rest of the terms in the action $S$ and from the Dirac-Born-Infeld action of extended objects wrapping internal cycles on $\mathbb{X}_6$, as D3-branes and orientifold planes $O3^-$.\\

However,  in the presence of orientifold planes, it is natural and expected to have torsional cycles. For instance, in a IIB toroidal orientifold, the quotient space $\mathbb{T}^6/\mathbb{Z}_2$ contains torsional cycles of different dimension (dual to torsional fluxes), meaning that there are cycles that after wrapping them a certain number of times, one ends up with a subspace of $\mathbb{T}^6$ with boundary. Since we are considering the presence of orientifold planes, we shall assume the existence of torsional cycles in  generic K\"ahler manifolds. Under this context we shall study whether or not $\omega_5$ contributes to $\eV$ via torsional cycles.\\

The $p$th-cohomology group of a six-dimensional K\"ahler manifold is written as
\begin{eqnarray}
\H^p(\mathbb{X}_6;\mathbb{R})&=&\H^p(\mathbb{X}_6;\mathbb{Z})+\text{Tor~}\H^p(\mathbb{X}_6;\mathbb{Z}),\nonumber\\
&=& \mathbb{Z}^{b_p} + \left(\mathbb{Z}_{k_1}\oplus \dots \oplus\mathbb{Z}_{k_n}\right),
\label{torsion}
\end{eqnarray}
where $b_p$ is the Betti number for $\H^p(\mathbb{X}_6,\mathbb{Z})$ and $k_i \in \mathbb{Z}$. Let us consider the case for $p=3$. A 3-form in the torsional part can be decomposed as
\begin{equation}
\pi^{\text{tor}}_3= \lambda^i \pi^{\text{tor}}_{3,i},
\end{equation}
with $i=1,\dots ,n$ according to (\ref{torsion}) and $\lambda^i\in \mathbb{Z}$.  In the case in which the set of integers $\lambda^i$ has a greatest common  divisor (gcd) $\kappa$,  there exists 
  a non-closed 2-form $\hat{\omega}_2$ such that $ d\hat{\omega}_2=\kappa\pi_3^{\text{tor}}$, i.e., $\pi_3^{\text{tor}}\in\mathbb{Z}_k$. The set of such 2-forms  is denoted $\hat{\Omega}^2(\mathbb{X}_6)$.  If $\lambda^i=\kappa^i k^i$ only for some $i$, then there exists $\hat{\omega}_i\in \hat{\Omega}_i^2(\mathbb{X}_6,\mathbb{Z})$ such that $d\hat{\omega}_i=k_i\pi_{3,i}^{\text{tor}}$. In this scenario, generic  RR and NS-NS potentials are given by
\begin{eqnarray}
C_2&=&c^a\omega_a+\tilde{c}^i\hat{\omega}_i,\nonumber\\
B_2&=&b^a\omega_a+\tilde{b}^i\hat{\omega}_i
\end{eqnarray}
where $\omega_a\in\H^2_-(\mathbb{X}_6,\mathbb{Z})$, $\hat{\omega}_i\in\hat{\Omega}^2(\mathbb{X}_6,\mathbb{Z})$ with $a=1\dots h^{1,1}_-(\mathbb{X}_6)$ and $i=1,\dots n$. The presence of 2-forms $\hat{\omega}_i$ implies that  the K\"ahler form $J_2$ can also be written as
\begin{equation}
J_2=t^a\omega_a+\tilde{t}^i\hat{\omega}_i,
\end{equation}
from which $dJ_2=k_i\tilde{t}^i\pi_{3,i}^{\text{tor}}$. Hence for $\tilde{t}^i=\tau^i/k_i$, $dJ_2$ is non trivial in $\H^3(\mathbb{X}_6,\mathbb{Z})$ and $\mathbb{X}_6$ is not a CY manifold but at least a K\"ahler manifold modulo $k_i$.\\

If now we restrict the compactification over a K\"ahler manifold with torsion as above, the contribution from $S_{F_5}$ is not longer zero, but
\begin{equation}
S_{F_5}=\frac{5!}{16\kappa_{10}^2}\int d\text{Vol}_4\int \left(H_3\tilde{c}^i-F_3\tilde{b}^i\right)\wedge\hat{\omega}_i\wedge de^{4A(y)},
\end{equation}
with $A(y)$ the warping factor in Eq.(\ref{metric}). Therefore, the contribution of $F_5$-form to the scalar potential, in the Einstein frame, is given by
\begin{equation}
V_{F_5}\sim \frac{A_5}{\tau^4},
\end{equation}
where $A_5= A ~\text{mod}~k_i$ for some $A$.

\subsection{Conditions for finding stable AdS vacua}
The above contribution to $\eV$ from $S_{F_5}$ together with the contributions from 3-form fluxes, D3-branes and $O3^-$-planes, lead us to a scalar potential of the form
\begin{equation}
\eV = \frac{A_{H_3} s}{\tau^3}+\frac{A_{F_3}}{s\tau^3}+\frac{A_{F_5}}{\tau^4}+\frac{A_{3}\mathcal{N}_{3}}{\tau^3},
\end{equation}
which actually has some stable AdS minima if there is at least one negative contribution from the above terms. However, since the flux contribution $A_{G_3}$ is positive definite\footnote{According to our previous analysis, this means that supersymmetry is broken by the dilaton modulus.} and $A_{F_5}$ is defined modulo an integer, the only option left  is that, from the contribution of 3-dimensional sources, $\mathcal{N}_3$ must be negative.\\

\begin{figure}[htbp]
   \centering
   \includegraphics[scale=0.2]{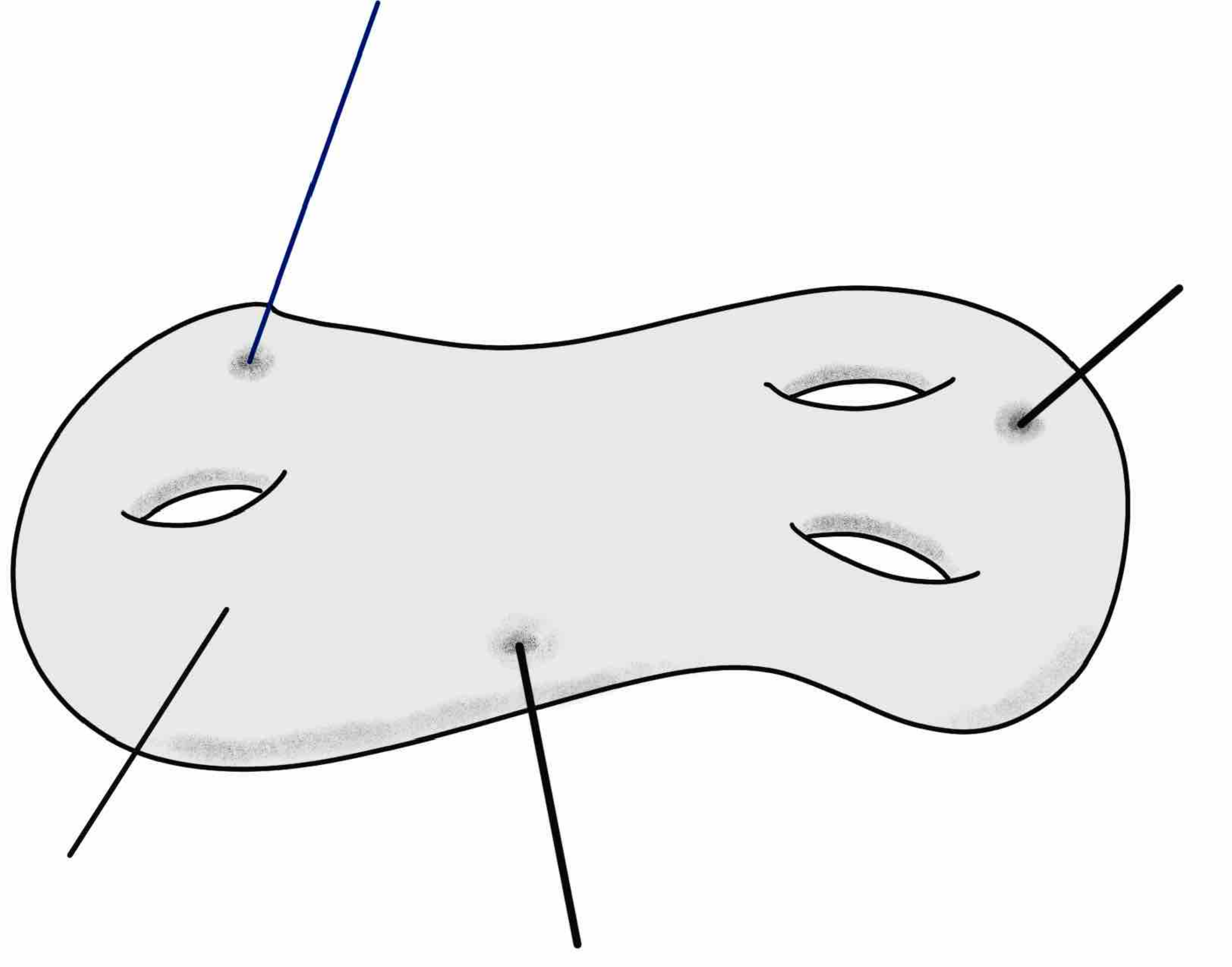} % requires the graphicx package
     \begin{picture}(0,0) 
  \put(-120,132){O3}
   \put(-90,0){O3}
   \put(-10,100){O3}
   \put(-193,7){D3 -O3}
     \end{picture}
   \caption{Schematic picture of the compact space and the locii of the orientifold planes and $D3$-branes.}
   \label{fig:D3$-$O3}
\end{figure}
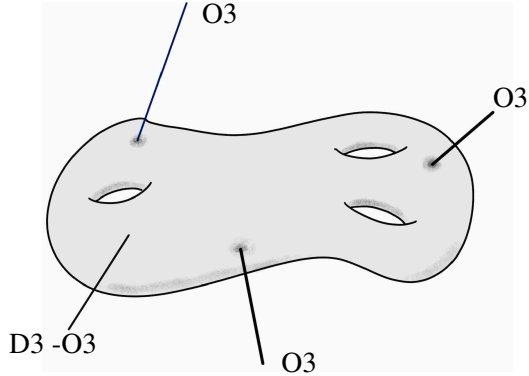

By restricting the flux configurations and local sources to satisfy that $\mathcal{N}_3<0$, the number of $O3^-$-planes must be larger than the number of $D3$-branes,  implying that at some points in the internal space, there must be isolated orientifold planes, or in other words  that there are no $D3$-branes of top of some of the $O3^-$-planes. This follows from the usual assumption that orientifold planes are immovable and from  the fact that there is an attraction between $D3$-branes and $O3^-$-planes due to the RR $D3$-brane charge they carry. For instance, the most simple configuration involving the presence of $D3$-branes with $\mathcal{N}_3<0$ is to  have 4 orientifold fixed points and a single  $D3$-brane sitting at one of those points. In such case, $\mathcal{N}_3=-1$ (see Figure \ref{fig:D3$-$O3} for a schematic representation of this configuration).

\begin{figure}[htbp]
   \centering
 a)\includegraphics[scale=0.5]{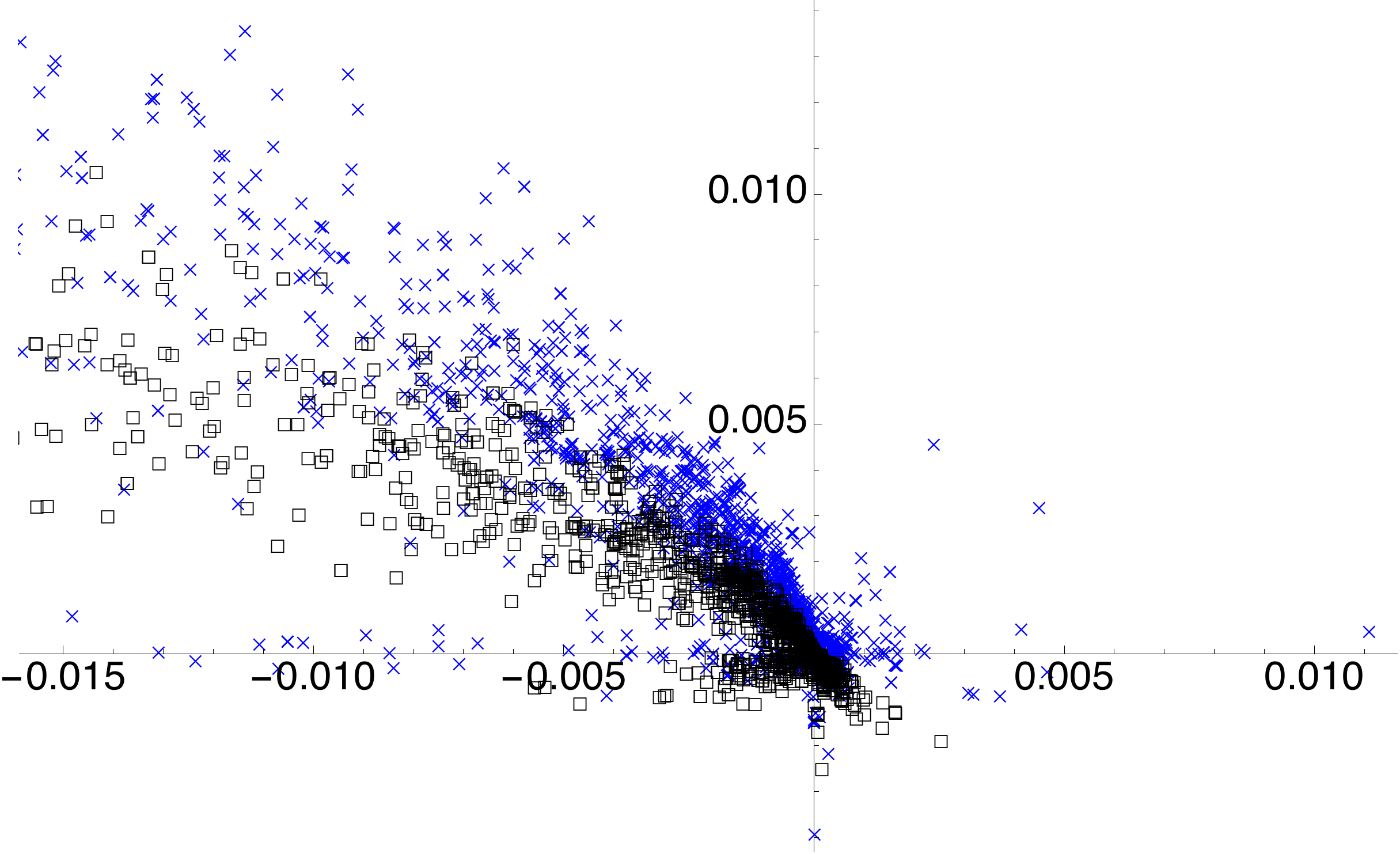}   
  \begin{picture}(0,0) 
  \put(-150,215){V$_0$}
   \put(-5,55){min $m^2$}
     \end{picture}\\
b) \includegraphics[scale=0.5]{ 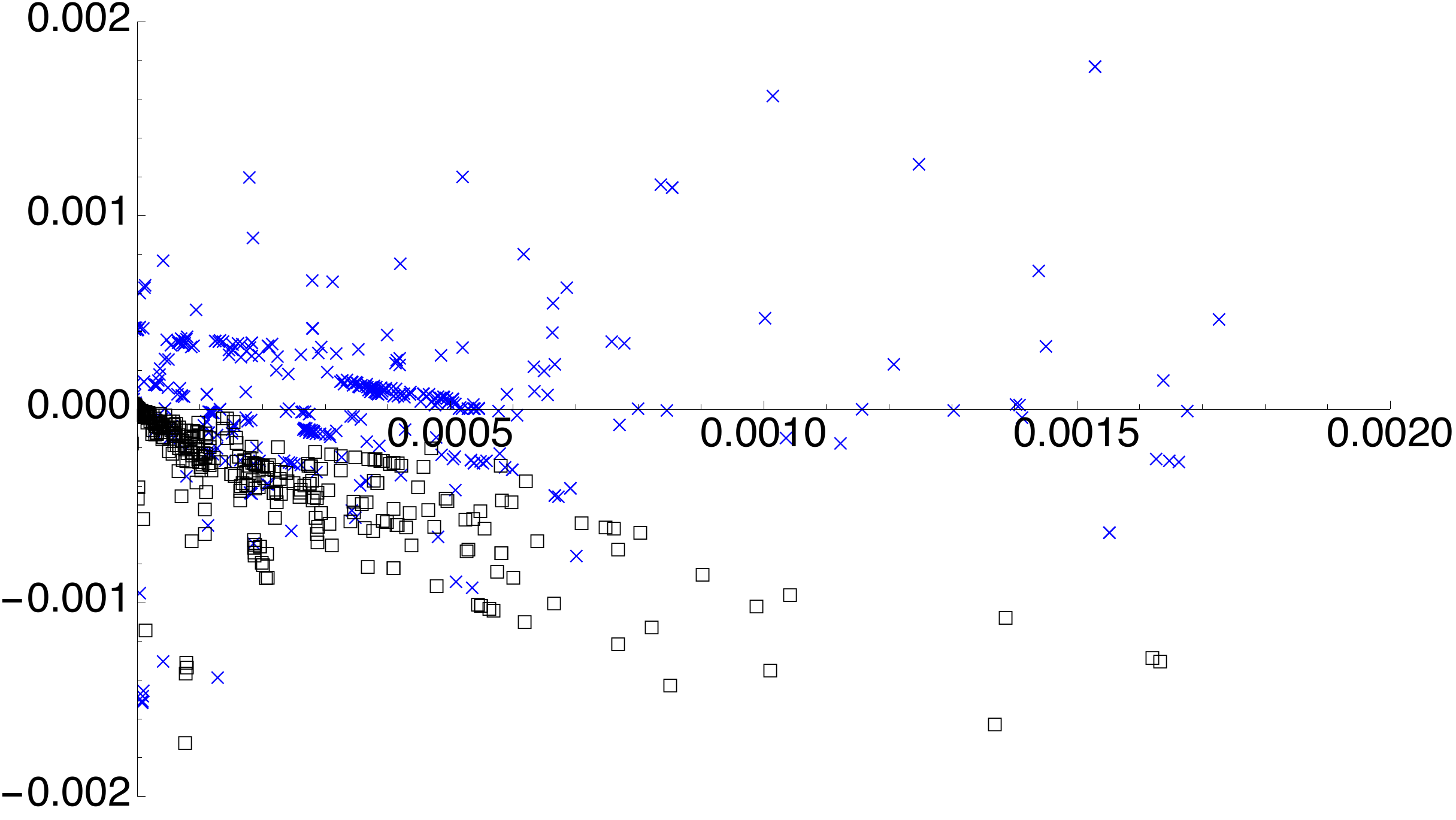}% requires the graphicx package
 \begin{picture}(0,0) 
   \put(-320,195){V$_0$}
   \put(-20,105){min $m^2$}
  \end{picture}
   \caption{Plot of the critical points found by the hybrid algorithm. The black squares correspond to the cases where  $F_5$ contributions were taken into account without non-BPS states, whereas the blue crosses consider the presence of $\hat{D5}$ non-BPS states. On the second image, we present a zoom of the stable cases.}
   \label{fig:landsbpsnonbps}
\end{figure}
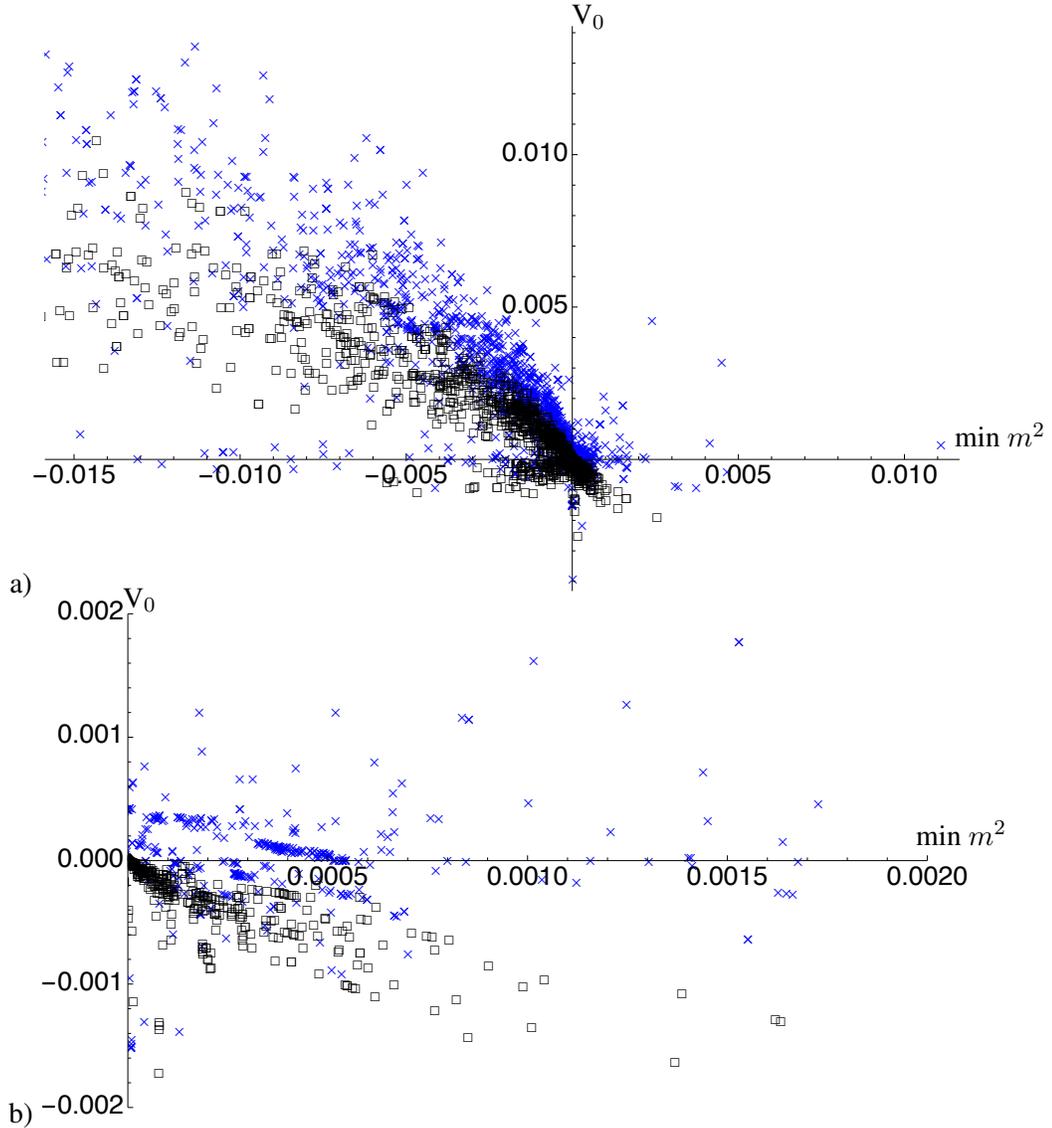

Under these conditions we implemented our ML algorithm described in Appendix \ref{ML}.  With it, we were able to find 389 different stable  AdS vacua. However, in spite of designing  our algorithm such that finding dS vacua was favored over AdS, no dS one was found. Our results are shown in Figure \ref{fig:landsbpsnonbps} where all found vacua, stable or not, are represented by black squares. \\

%%%%%%%%%%%%%%%%%%%%%%%%%%%%%%%%%%%%%%%%%%%%%%%%%%%%%%%%%%%
%%%%%%%%%%%%%%%%%%%%%%%%%%%%%%%%%%%%%%%%%%%%%%%%%%%%%%%%%%%

\section{Stable dS vacua from Non-BPS states}

The presence of torsion  opens up the possibility to consider wrapping $D$-branes on torsional cycles.  The existence of torsional cycles follows from the dual maps between homology and cohomology, where
\begin{equation}
\int_{\Sigma^{j,\text{tor}}_2}\hat{\omega}_i=\int_{\mathbb{X}_6}\hat{\omega}_i\wedge PD(\Sigma^{j,\text{tor}}_2)=\delta^j_i.
\label{2cycleform}
\end{equation}
with
\begin{equation}
k_i\Sigma^{i,\text{tor}}_2=\partial\hat{\Pi}_{3}^i.
\end{equation}
This last assertion means that the homology group $\text{H}_2(\mathbb{X}_6, \mathbb{R})$ also has a torsion component, i.e.,  $ \Sigma^{i,\text{tor}}_2\in \text{Tor~}\H_2(\mathbb{X}_6, \mathbb{Z})$ and $\hat{\Pi}_{3}^i\in \hat{\Omega}_3(\mathbb{X}_6,\mathbb{Z})$. It follows then that $\text{Tor~}{\H}_2(\mathbb{X}_6,\mathbb{Z})\sim \text{Tor~}{\H}^3(\mathbb{X}_6,\mathbb{Z})$. We shall follow the argument in which these  states $-$D-branes wrapped on torsional cycles$-$ are in fact related to the well-known non-BPS states constructed from K-theory \cite{Witten:1998cd}.\\

The existence of non-BPS states in the presence of an orientifold plane $O3^-$ can be inferred by applying T-duality on the corresponding coordinates on a torus compactification of Type I string theory, which actually have non-BPS branes as $\hat{D7}$, $\hat{D8}$, $\hat{D0}$ and $\hat{D(-1)}$. Hence, by taking for instance a non-BPS $\hat{D7}$-brane spanned on 4 coordinates on $T^6$ immersed in an $O9^-$-plane and applying T-duality on the compact coordinates, we get an extended $O3^-$-plane and a 5-brane wrapping a 2-dimensional space in the covering space. We expect this object to carry a topological $\mathbb{Z}_2$ charge as its T-dual partner. Indeed, by computing the 2nd-homology group of $\mathbb{T}^6/\mathbb{Z}_2$ we see that there are torsional 2-cycles. Wrapping $D5$-branes of type IIB theory on such cycles seem to be the way to construct the aforementioned non-BPS states. Moreover,  by computing the corresponding T-dual K-theory group one sees that stable non-BPS states are present, carrying discrete topological charge $\mathbb{Z}_2$ with three extended coordinates while the other are wrapped on the compact space.\\

For a more general compactification, one must compute the K-theory groups of intersecting sources, i.e., of configurations of branes intersecting orientifold planes wrapping cycles on a compact manifold. This is indeed a difficult task. However, ignoring the compact component of the space, it is possible to classify intersecting branes with orientifolds by the use of the Kasparov KK-theory \cite{Asakawa:2002nv, Garcia-Compean:2008fzo}. Since the formulation is quite technical and it is beyond the scope of this work\footnote{The KK-theory group  classifying $Dd$-branes on top of an $Op^-$-plane, with $p=3 ~\text{mod}~ 4$ and $d>p$ is given by \cite{Garcia-Compean:2008fzo}
\begin{equation}
KKH^{-2}(\mathbb{R}^{d-s,r}, \mathbb{R}^{9-p, p+r-s})=KO(\mathbb{S}^{2p-2s+d-3}),
\end{equation}
where $s$ are the number of coordinates of the D$d$-brane overlapping the orientifold plane and $r$ is the codimension of the $Dd$-brane inside the orientifold. For a $D5$-brane on top of an $O3^-$-plane with 2 transversal coordinates, $p=s=3$, $r=0$ and $d=5$.}, we just present the KK-theory group which classifies 5-branes fully intersecting an $O3^-$-plane, i.e., with 2 transversal coordinates to the orientifold plane and its relation to orthogonal K-theory group. This is:
\begin{equation}
KKH^{-2}(\mathbb{R}^{2,0}, \mathbb{R}^{6,0})=KO(\mathbb{S}^2)=\mathbb{Z}_2,
\end{equation}
as expected.\\

Based on these results we are taking as valid the construction of stable non-BPS states by wrapping D-branes on torsional cycles of a K\"ahler manifold $\mathbb{X}_6$. In particular, we can construct a non-BPS $\hat{D5}$-brane by wrapping a $D5$-brane  on a torsional 2-cycle $\Sigma_2^{\text{tor}}\in \H^{\text{tor}}_2(\mathbb{X}_6, \mathbb{Z})$, where $\Sigma^{\text{tor}}_2$ is the cycle where the 2-form $\hat{\omega}_2$ is supported as in Eq.(\ref{2cycleform}).\\

Summarizing, a compactification on a K\"ahler manifold $\mathbb{X}_6$ with torsional components in (co)-homology, leads us to the possibility to include D-branes wrapping torsional cycles.  Here we shall consider the contribution to the effective scalar potential from non-BPS $\hat{D5}$-branes. However, before that we must discuss possible sources of instability on a configuration constructed with  fluxes, $D3$-branes, $O3^-$-planes and non-BPS states.\\

\subsection{Consistency by adding non-BPS $\hat{D5}$-branes}
As it is known \cite{Witten:1998cd}, the non-BPS brane $\hat{D7}$ in type I theory can be constructed by a pair of a $D7$ and $\bar{D7}$-branes, where the tachyon on the open sector string connecting  the two branes is projected out by the orientifold $O9^-$.  However, since in type I theory there are 32 $D9$-branes, there is also a tachyon from the open string between $D9$-branes and $D7$-branes, making the non-BPS $\hat{D7}$-brane to be unstable \cite{Loaiza-Brito:2001yer}.\\

In a T-dual version, upon compactification on a six-dimensional torus,  the above configuration is mapped into $D3$-branes and $O3^-$-planes sitting at different points on $\mathbb{T}^6$ and $D5$-branes wrapping torsional 2-cycles on the compact space, corresponding to the non-BPS states $\hat{D5}$.
Therefore, by T-duality, it is expected that in a given fixed point in the internal space, a $\hat{D5}$-brane  coinciding with at least one $D3$-brane, would be unstable to decay into a field configuration while preserving its topological charge $\mathbb{Z}_2$ . This instability is not present (at least locally) if at the given fixed point, there are not $D3$-branes, a configuration we can have if there are more  orientifolds than $D3$-branes, i.e., if $\mathcal{N}_3=\mathcal{N}_{D3}-\frac{1}{2}\mathcal{N}_{O3}<0$. In order to cancel the $D3$-brane charge tadpole, we then require a positive contribution from fluxes. These two characteristics, ${\cal N}_3<0$ and $\mathcal{N}_{\text{flux}}>0$ are essential to guarantee the stability of adding non-BPS $\hat{D5}$-branes. Notice that $\mathcal{N}_3<0$ is one of the conditions to assure the existence of stable AdS vacua without adding non-BPS states.\\

Under the above circumstances, we shall take a $D5$-brane and wrap it on a torsional 2-cycle $\Sigma^{\text{tor}}_2\in \text{Tor}~\H_2(\mathbb{X}_6, \mathbb{Z})$. Following \cite{Bergman:2001rp}, we argue that such a state is classified by the corresponding K-theory group on $\mathbb{X}_6$. Also, we shall consider the contribution of this non-BPS  $\hat{D5}$-brane to the effective scalar from the DBI term. However, it is important to notice that its contribution must be measured as $\text{mod} ~2$ since a pair of non-BPS branes with topological charge $\mathbb{Z}_2$ anihillate to each other. This means that if the total discrete charge vanishes, the effective contribution from non-BPS branes is null \cite{Blumenhagen:2019kqm, Blumenhagen:2021nmi}. Another important fact we must have in mind is that we are ignoring torsional components for 3-form fluxes, although there is no restriction for their presence\footnote{In \cite{Damian:2019bkb} some consequences of turning on torsion components of fluxes are commented.}. \\

Hence, the effective contribution of a non-BPS brane $\hat{D5}$ at leading order in $\alpha'$ is given by the DBI action, 
\begin{equation}
S_{\hat{D5}}=-2T_5\int d^6\xi ~e^{-\phi} \sqrt{-\widetilde{g}_6}
\end{equation}
where $\widetilde{g}_6$ is the determinant of the induced metric on the $\hat{D5}$-brane worldvolume. Therefore, the corresponding effective scalar potential in the Einstein frame reads
\begin{equation}
V_{\hat{D5}}=\frac{{A}_{\hat{D5}}}{s^{1/2}\tau^{5/2}},
\label{VD5}
\end{equation}
where $2n{A}_{\hat{D5}}=0$ for $n\in \mathbb{Z}$.\\

\subsection{Stable dS vacua with non-BPS states}
 In order to look for dS minima we shall employ an hybrid method which consists in applying a stochastic method known as Simulated Snnealing followed by the gradient descent algorithm (see Appendix \ref{ML}). The effective scalar potential constructed by contributions from 3-form fluxes, 3-dimensional sources, a torsional component of $F_5$ and non-BPS $\hat{D5}$-branes is
\begin{equation}
\eV = \frac{A_{H_3} s}{\tau^3}+\frac{A_{F_3}}{s\tau^3}+\frac{A_{F_5}}{\tau^4}+\frac{A_{3}\mathcal{N}_3}{\tau^3}+ \frac{A_{\hat D5}}{{s}^{1/2}\tau^{5/2}}.
\end{equation}
As discussed in \cite{Shiu:2011zt} (see also \cite{Hertzberg:2007wc}), it is expected that this anzats evades the no-go theorems and increases the possibility to find some stable dS vacua. \\
%The A$_{\hat D5}$ is related to the stable non-BPS state and its contribution is equal in magnitude to a O5 plane. The case in which this term is related to O5 planes implies that A$_{\hat D5} < 0$. 

In Figure \ref{fig:landsbpsnonbps} it is shown by blue crosses, the critical points found by the above-mentioned algorithm. Notice the presence of many stable dS vacua. In Table \ref{tab:dSv} we present the explicit values of  the scalar potential contributions for some of these vacua.

\begin{table}[!ht]
\centering
$
\begin{array}{|c c c c c c c c c c|}
\hline
 \text{min}\,V  & m_{\tau}^2 & m_{\text{s}}^2 & \tau & \text{s}  & \text{A}_{F_3} & \text{A}_{H_3} & \text{A}_{F_5} & \text{A}_{3}\mathcal{N}_3 & \text{A}_{\hat D 5} \\
 \hline
 1.309 \times 10^{-6} & 0.000546 		& 0.003728 	& 3.695 	& 2.363 		& 0.7628		& 0.2486 		& 0.9769 	& -1.704 	& 0.4231 \\
 5.676 \times 10^{-6} & 0.0005166 		& 0.003874	& 3.727	& 2.298		& 0.7566		& 0.2575		& 0.9755	& -1.708	& 0.4125 \\
6.980 \times 10^{-5} & 0.0006562		& 0.007878	& 2.917	& 2.111		& 0.7463		& 0.2189		& 0.3022	& -1.135. &  0.1851 \\
9.561 \times 10^{-5} 	& 0.0003757		& 0.003277	& 3.855	& 2.373		& 0.7638		& 0.2495		& 0.9778	& -1.702	& 0.4238 \\ 
4.039 \times 10^{-4}	& 9.460 \times10^{-7}& 0.004265	&4.258	&  2.170 		& 1.260		& 0.3864		& 0.6982	& -2.067	& 0.3677 \\
 \hline
 \end{array}
 $
 \label{tab:dSv}
 \caption{Selected vacua found by the hybrid SA+CG algorithm.}
\end{table}

\section{Uplifting conditions by non-BPS states}
In this section we are interested in discussing the uplifiting of AdS  stable vacua to dS by the presence of non-BPS states as the $\hat{D5}$-branes. As previously observed, a dimensional reduction in the presence of 3-form fluxes $H_3$ and $F_3$, as well as 3-dimensional sources as $D3$-branes and $O3^-$-planes together with a torsional $F_5$ form, leads us to the possibility to construct AdS stable vacua. 
%In this section we want to describe the conditions upon which stable AdS vacua, are uplifted to apparently stable dS vacua.
For $A_{D5}=0$, the minimum for $\eV$ is located at 
%the zeros of a polynomial  with formal derivative not equal to zero in the field extension $\mathbb{Q}\left( \left(A_{F_3}/A_{H3} \right)^{1/2} \right)$ which belongs to the Galois group $\mathbb{Z}/2\mathbb{Z}$, where the physical solutions ($s, \tau > 0$ ) take the values 
\eq{
s_0= \left( \frac{A_{F_3}}{A_{H_3}} \right)^{1/2}  \quad \tau_0 = \frac{4}{3} \frac{A_{F_5}}{\Delta}
\label{vevsAdS}
}
for $\Delta = -(A_3 \mathcal{N}_3+2 A_{H_3}^{1/2} A_{F3}^{1/2})$.  Notice that in the case we are turning on a single flux $G_3$, meaning that we are considering a contribution to the superpotential along one single period, $\Delta$ reduces to zero due to the tadpole cancellation. Therefore, it is necessary to consider more than one flux in order to uplift the AdS vacua while keeping $|A_3\mathcal{N}_3|>2(A_{H_3}A_{F_3})^{1/2}$ such that $\tau_0>0$. Therefore we require that two specific conditions must be taken:
\begin{enumerate}
\item $W=\int G_3\wedge \Omega$ must be constructed from more than just one period.
\item $\mathcal{N}_{O3}>4\frac{(A_{H_3}A_{F_3})^{1/2}}{A_3}+2\mathcal{N}_{D3}$.
\end{enumerate}

We shall restrict the rest of our analysis to such a case.\\

The minima of the AdS can be written in function of the vacuum expectation value (vev) of the K\"ahler modulus as
\eq{
{\mathcal{V}}_{AdS} = - \frac{1}{3}\frac{A_{F_5}}{\tau_0^{4}} \,,
}
thus, the larger $\tau_0$, the smaller value for the AdS vacua, which  is compatible with the KKLT scenario. The eigenvalues can be written in terms of the vev's as
\eq{
m^2_s = \frac{2 A_{H_3}^{1/2}}{s_0 \tau_0} \quad \text{and} \quad m^2_{\tau} = \frac{4 A_{F_5}}{\tau_0^6} \,,
}
and we see that for large values of $\tau_0$,  the smallest eigenvalue is always in the $\tau$ direction. \\

Now, to uplift from stable AdS to  dS vacua it is necessary to add energy associated to the non-BPS states $\hat{D5}$ as in Eq.(\ref{VD5}), which change the vevs of the moduli shifting its numerical values to greater values. In this case  the K\"ahler modulus modify to
\eq{
\tau = \frac{4(-A_{F_3}+ A_{H_3}s^2)^2}{A_{D5}^2 s}\,,
}
which in the limit of $A_{\hat{D5}}\ll 1$ can be written as 
\eq{
s &= s_0 + \frac{1}{2\sqrt{3}}\left( \frac{A_{F_5}}{A_{F_3}^{1/2} A_{H_3}^{3/2}} \right)^{1/2} \frac{1}{( \Delta )^{1/2}} A_{\hat{D5}}+ \mathcal{O} \left( A_{\hat {D5}}^2 \right) \, \\
\tau &=\tau_0 + \frac{1}{2^6}\frac{\tau_0^2}{A_{F_3}^{3/2} A_{H_3}^{1/2}}A_{\hat{D5}} +  \mathcal{O} \left( A_{\hat {D5}}^2 \right) \,.
}
Notice from this and from Eq.(\ref{vevsAdS}) that for $\Delta > 0$ this branch of solution takes real values. In this context, one also can express the effective potential at leading terms in $A_{\hat{D5}}$ as
\eq{
\eV = V_{AdS} + \frac{1}{s_0^{1/2} \tau_0^{5/2}} A_{\hat {D5}} +  \mathcal{O} \left( A_{\hat {D5}}^2 \right)
}
where the uplifting from AdS  to dS depends on how deep is the AdS vacuum. \\

 However it is important to analyze whether the uplifting would be stable or not. For that we shall study under which conditions there are tachyons. Let us start by establishing the required stability criteria for the AdS vacua.  Since we are interested only in their presence, we shall take the the mass matrix as
\begin{equation}
(M^2_{AdS})_{ij}=\partial_{ij}\mathcal{V}_{\text{AdS}},
\end{equation}
with $i,j=s, \tau$. The eigenvalues $\lambda_{AdS}$ are given by
\eq{
 \lambda_{AdS} = \frac{1}{2}\tr M^2_{AdS} \pm \alpha
}
%Now, let us consider the uplift from AdS to dS. This shall be done by taking $A_{\hat{D}5}= 0$ and calculate the minima with all the other factors equal. This is by keeping the $A_{H_3}$, $A_{F_3}$, $A_{F_5}$ constant (for this examples we shall take $A_{R_6} = 0$ since the contribution coming from the curvature of the internal space is not required to obtain the stable dS). In this cases, it is posible to find a stable AdS and by adding the $A_{\hat D 5}$ the uplift is obtained by keeping the stability of the vacua. In this way this is manner to contruct dS vacua from a AdS provided that the AdS is free of tachyons.\\
where  $\alpha =  \sqrt{(\tr M^2_{AdS})^2- 4 \det M^2_{AdS}}$. According to the Silverster's criterium, a stable minimum exists always that $\tr M^2_{AdS} > 0$ and $\alpha$ be real. Notice that large values for the eigenvalues $\lambda_{AdS}$ indicate that it is difficult to destabilize the minimum. On the contrary, small values of $\lambda_{AdS}$  correspond to  very flat potentials from which it is easy to  escape from.  Following this line of reasoning, we want to show that by adding non-BPS states $\hat{D5}$ the eigenvalues related to an AdS vacuum become smaller.\\

For that,  let us consider adding the contribution from non-BPS states $\mathcal{V}_{\hat{D5}}$, such that
\begin{equation}
(M^2)_{ij}=\partial_{ij}\left(\mathcal{V}_{AdS}+\mathcal{V}_{\hat{D5}}\right)=(M_{AdS}^2)_{ij}+(M_{\hat{D5}})_{ij}
\end{equation}

One realizes that the eigenvalues for each of the moduli decreases as we add the $A_{\hat D5}$ term. To clearly show this, lets us split   $\tr M^2$ and $\det M^2$ in terms of the contributions of $A_{\hat D5}$ as
 \eq{
 \tr M^2 = \tr M^2_{AdS} + f(A_{\hat D5}), \qquad \det M^2 = \det M^2 _{AdS}+ g(A_{\hat{D5}}) \,, 
 }
 where $f(A_{\hat D5})$ and $g(A_{\hat D5})$ are positive definite homogeneous functions of degree 1 on $A_{\hat D5}$. If the added potential is of the form
 \begin{equation}
 \mathcal{V}_{\hat{D5}}\sim \frac{1}{s^m\tau^n},
 \end{equation}
 with $n,m>0$, which indeed is our case. Thus, by adding the $A_{\hat D5}$ terms, there is a contribution $\delta\lambda$ to the eigenvalues $\lambda_{AdS}$ as
 \eq{
 \lambda = \lambda_{AdS}+ \delta \lambda.
 }
 In this context, we say that if $\delta \lambda < 0$, the eigenvalues of the mass matrix decrease due to the contribution of the non-BPS states. Indeed, the change of the eigenvalues can be written explicitly as
\eq{\label{eigenvalues}
\delta \lambda = \frac{1}{2}(f\pm\alpha) \left( 1 - \sqrt{1+ \gamma} \right)
}
where 
\eq{
\gamma = \frac{2 f \left( \tr M^2_{AdS} \pm \alpha \right) + 4 g}{(f\mp \alpha)^2}.
}
Since $f$ and $g$ are positive functions and $\alpha < \tr M^2$, then $\gamma$ is positive definite. In consequence the term $\left( 1 - \sqrt{1+ \gamma} \right)$ shall be negative. This in general implies that
\eq{
\frac{2(\delta \lambda)}{(f\pm \alpha)} \leq 0.
}
Finally, putting $f$ and $\alpha$ in terms of the determinant and trace of the mass matrix we find that
\eq{
f\pm \alpha = \tr M - \tr M^2_{AdS} \pm \sqrt{(\tr M^2_{AdS})^2 - 4\det M^2_{AdS}} > 0,
}
and $\delta \lambda < 0$. \\

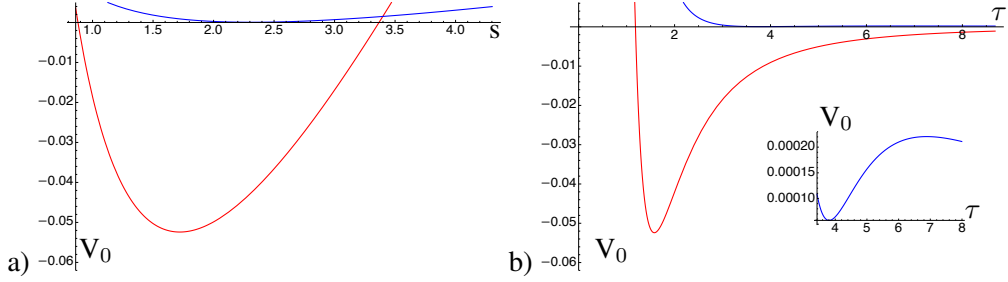
\begin{figure}[htbp]
   \centering
   a) \includegraphics[scale=0.25]{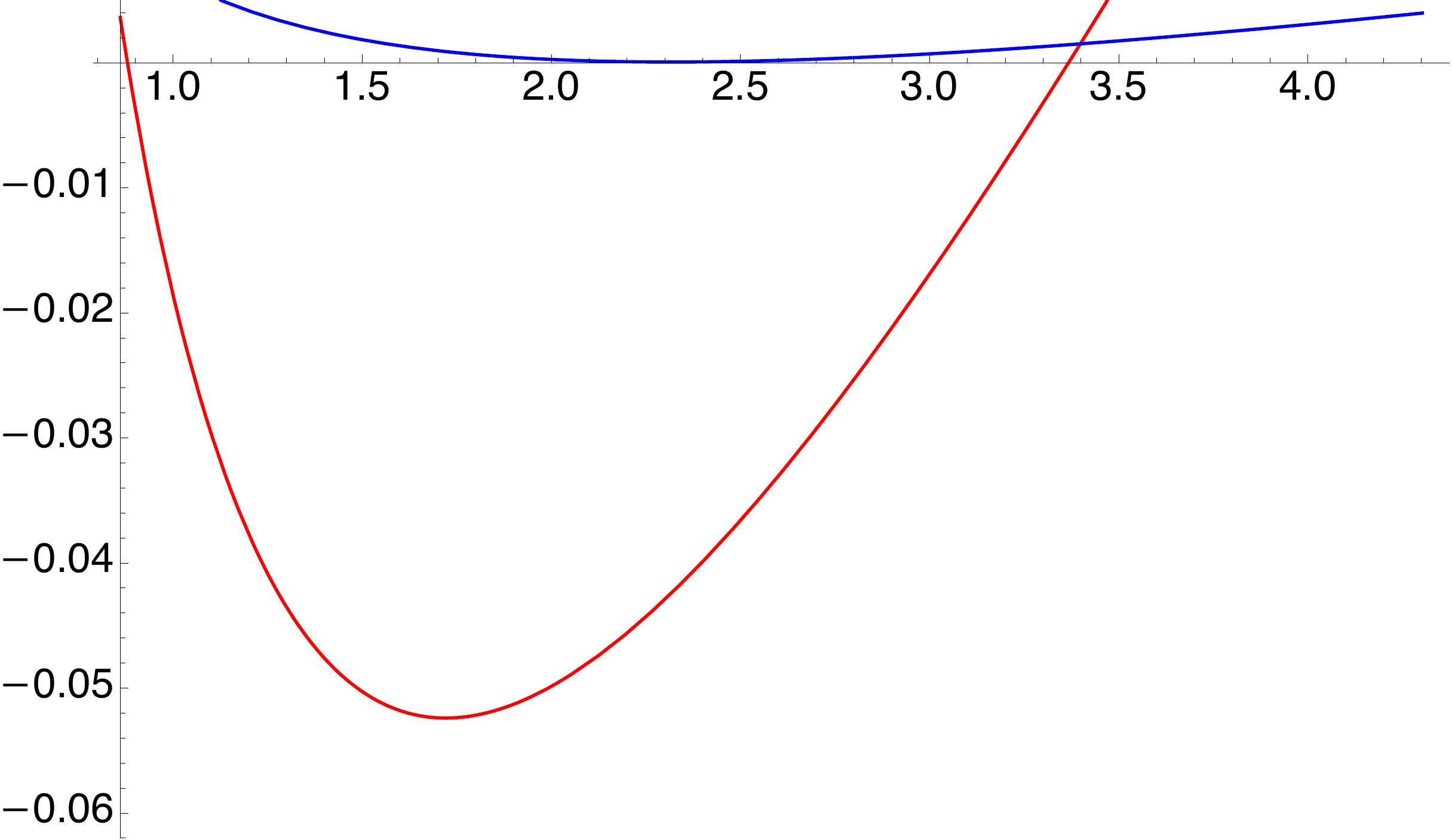} b)  \includegraphics[scale=0.25]{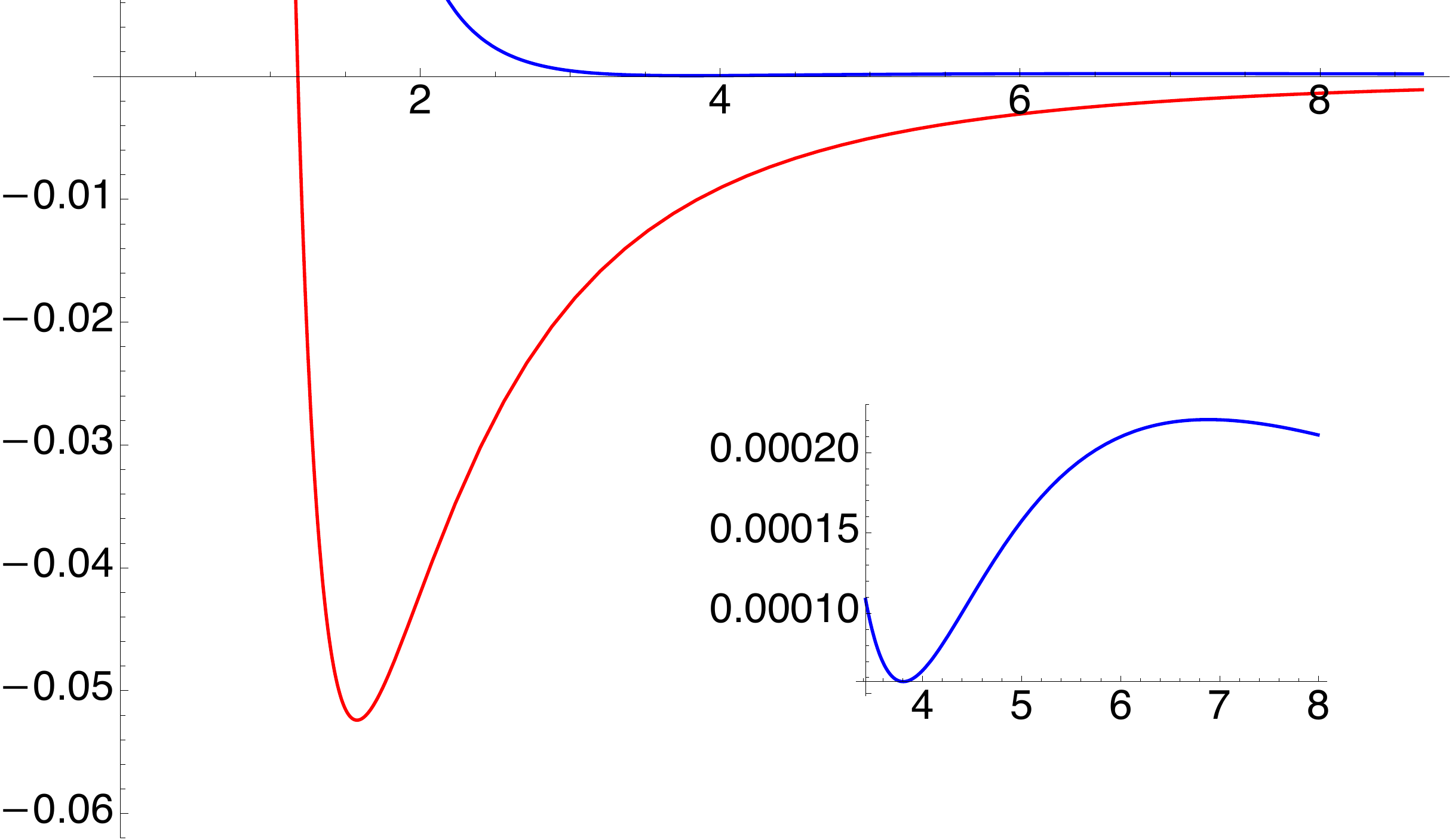}% requires the graphicx package
      \begin{picture}(0,0) 
  \put(-350,5){V$_0$}
   \put(-195,88){s}
     \put(-155,5){V$_0$}
   \put(-5,95){$\tau$}
    \put(-70,55){V$_0$}
   \put(-15,20){$\tau$}

     \end{picture}
   \caption{Plots for the uplift mechanism by employing the A$_{\hat D 5}$ contribution. In red we show the effective potential in an AdS minimum defined by the contributions $A_{F_3}= 0.77046$, $A_{H_3}= 0.24018$, $A_{O_3}=-1.6974$, $A_{F_5}=0.97955$ and with moduli vevs given by $s =1.7911$, $\tau = 1.5603$. By adding non-BPS $\hat{D5}$-branes with $A_{D_5}= 0.43564$, it is observed that  the uplift reduces the mass of the scalar field while its expectation value moves to the right as $s= 2.4473$, $\tau = 3.8434$. Notice that the uplift of the K\"ahler moduli produce a nearly flat direction, which is compatible with the KKLT scenario. }
   \label{fig:uplift}
\end{figure}

Adding non-BPS states drives two important features in the effective potential. In one hand, uplifts the value of $\mathcal{V}_{AdS}$  to a dS one, but in the other hand, since the contribution to the energy at the minimum is positive, the scalar potential becomes very flat increasing the probabilities for this vacuum to be destabilized.
We show this behavior, for one case,  in Figure \ref{fig:uplift}.\\

\subsection{Comments about some Swampland conjectures}
We have described a way to construct a dS vacuum by adding the contribution to the scalar potential from a non-BPS $\hat{D5}$-brane to  a non-SUSY AdS vacuum ($D_SW\ne 0$). However, as recently studied, there are some constraints around the construction of both states. First of all, it has been argued that a non-supersymmetric AdS vacuum is at most metastable in the context of the Swampland program \cite{Freivogel:2016qwc,Ooguri:2016pdq}. Second of all, it is expected a constraint on the AdS scale with respect to the lightest moduli mass, and finally, in case of uplifting the non-SUSY vacuum to a dS one, the final vacuum is at most, metastable. Let us comment about these three points and how they are manifested in our setup.\\

As mentioned, one way to assure the construction of an AdS vacuum by considering the contribution of $F_5$ in a manifold with torsion, implies the stabilization of the complex structure by $D_U\mathcal{W}=0$ while keeping $D_S\mathcal{W}\ne 0$. Therefore, the AdS vacuum is non-SUSY. According to the Swampland conjectures, such an AdS vacuum must be at most metastable.  In our case, the source for instabilities could come from two places: first, from our assumption of not considering torsional components of 3-form fluxes, which usually drives some topological transitions as pointed out in \cite{Damian:2019bkb}. Second, since the contribution from $F_5$ is based on the existence of torsional cycles, it is possible that the total discrete charge must vanish following the recent proposal about having zero global charges in Quantum Gravity and its relation to K-theory by cobordisms as proposed in \cite{Blumenhagen:2021nmi}. We believe that both aspects are in fact related.\\

The second point concerns the AdS scale which it is also conjectured to satisfy a relation of the form
\eq{
m_{\text{mod}} R_{\text{AdS}} \gg c'
}
where $c' \sim 1$ and $R_{\text{AdS}} \sim |V_0|^{1/2}$ in order to keep a robust realization of a dS vaccum. Recent studies argue that effective models which support such a parametric hierarchy are in fact in the Swampland.
%The AdS conjecture relates the scale of the AdS vacua to the lightest mass of the moduli. In general, it is argued that it is not posible to separate these two scales and as the AdS scale reduce the lightest state increase otherwise the vacua it is in tension with the WGC. Indeed, in order to no destabilize the vacua it is required that
%. Thus, in absence of non-BPS states the ratio of scales can be written as,
Again, in our case the above two factors can be expressed in terms of each of the contributions to the scalar potential, for which we obtain that
\eq{
m_{\text{mod}} R_{\text{AdS}} =  \frac{3\sqrt{3}}{2} \frac{2 A_{{H}_3}^{1/2}A_{{F}_3}^{1/2}+A_{3}\mathcal{N}_3}{A_{{F}_5}} .
\label{mRAdS}
}
As all the constants $A_i$ for $i = \{ {H}_3, {F}_3, D3, O3 \}$ are of the same order, the energy added by $F_5$, for a $\mathbb{Z}_k$ discrete torsion, vanishes up to a multiple of $k$. Hence, unless $k$ is too large, the quotient (\ref{mRAdS}) is slightly larger than  order 1, and by taking $k=2$, $m_{\text{mod}}R_{AdS}\gtrsim c'$.\\

In this context, it is possible to add energy for the uplifting in such a way we stay in a region where stability can be (parametrically) controlled. Indeed, in our model, the AdS vacua do not contain tachyons neither in the axio-dilaton nor along Kähler directions. Besides,  the scale of the AdS is smaller that the energy coming from the lightest moduli violating the AdS conjecture. Thus, by adding a non-BPS states which its energy contribution scales a $s^{-1/2} \tau^{-5/2}$ generates a flattering effect accordingly. \\

Finally, according to the Swampland conjectures, a source of instabilities is expected  to affect the uplifted dS vacuum.  They could  come from the fact that the pair $\hat D5-D3$ (dual to the $\hat D7$ -$D9$)  is unstable \cite{Frau:1999qs, Lerda:1999um} and although a decay into a final state does not dilute  the discrete charge,  it is canceled out by requiring a vanishing K-theory charge \cite{Uranga:2000xp, Loaiza-Brito:2001yer}. However, in our case $\hat{D5}$-branes come from $D5$-branes wrapping torsional 2-cycles around an $O3^-$-plane with no $D3$-branes. Hence, at least locally, there are no instabilities at such points.
%due to the negative value of $A_{3}$ there is an energy excess of contribution coming from the O3 planes, which is not related to the dual version of the $\hat D7 - $O9 case, since in this case we have $%\hat{D5}$-branes wrapping torsinoal two-cycles around an orientifold fixed point.   
Thus, the non-BPS states are stable  and the only decay channel is through tunneling leading to the decompactification limit \cite{Brown:2010mf} probably described by a topological transition driven by torsional 3-form fluxes, as suggested in \cite{Damian:2019bkb}. A detail study about this process is reserved for a future work.\\

\section{Conclusions and Final comments}
As expected, the incorporation of $F_5$ fluxes by contribution to the effective scalar potential $\eV$ seems to be fundamental to find classical stable dS vacua in an orientifolded flux compactification of string theory. However, since in a Calabi-Yau manifold $F_5$ does not contribute to $\eV$, we need to consider other internal manifolds, such as the considered in \cite{Candelas:2014jma,Candelas:2014kma}.\\

As shown in \cite{Cai:2014vua} a K\"ahler manifold admitting torsion is a suitable example in which $F_5$ contributes to $\eV$. Moreover, these type of manifolds allow to wrap $D5$-branes on torsional cycles, by which one can construct non-BPS states actually classified by K-theory, with a non-zero contribution to $\eV$.\\

Under these circumstances and by implementing a novel ML algorithm we were able to find more than 200 dS critical points for $\eV$ out of which 170 are stable.\\

We also find that there are certain specific conditions that our configurations of branes and fluxes must fulfill in order to generate a stable dS vacuum by uplifting an AdS one. First, to obtain a stable AdS it is necessary to turn on the torsional part of $F_5$ and to have a configuration of branes and orientifolds such that the number of $O3$-planes or fixed points are larger than the number of $D3$-branes implying that $\mathcal{N}_{flux}>0$. Second, for these vacua to be uplifted to dS by incorporating the non-BPS states $\hat{D5}$ we ougth to have that:
\begin{enumerate}
\item The RR and NS-NS 3-form fluxes are supported in more than a single cycle,
\item $\mathcal{N}_{O3}>4\frac{\sqrt{A_{H_3}A_{F_3}}}{A_3}+2\mathcal{N}_{flux}$,
\end{enumerate}
where $A_{H_3}$ and $A_{F_3}$ are the contributions to $\eV$ (independent of moduli) from the fluxes $H_3$ and $F_3$, while $\mathcal{N}_3 A_3$ is the corresponding from 3-dimensional sources, with $\mathcal{N}_3=\mathcal{N}_{D3}-\frac{1}{2}\mathcal{N}_{O3}$. Under these conditions, it is possible to obtain that all mass eigenvalues are positive under the uplifting by non-BPS states. We observe, that 
\begin{enumerate}
\item
Getting a small positive value for $V_{\text{min}}$ seems to be a natural consequence by uplifting AdS vacua with a small deep. There are two consequences of this: the resulting uplifted potential is very flat while the probability for a destabilization of the dS vacua increases since at the limit for large volume, the potential goes to zero, indicating the presence of a barrier potential between the dS local vacuum and the the run away region for the K\"ahler moduli.
\item
We believe that this possibility of the scalar potential to become unstable could be generated by  extra mechanisms or topological transitions driven by torsional components on the 3-form fluxes as suggested in \cite{Damian:2019bkb} and in consequence, establishing a rich scenario where Swampland conjectures can be tasted.\\

\end{enumerate}

\begin{center}
{\bf Acknowledgments}\\
\end{center}

We thank to S. Ledesma-Orozco for its earlier collaboration. O. L.-B. thanks Nana Cabo, Yessenia Olguin and Ivonne Zavala for very nice discussions about related topics. C.D. is supported by CIIC-UG-DAIP No. 126/2022.  O. L-B is supported by CONACyT project CB-2015,  No. 258982 and by CIIC/UG/DAIP  No. 236/2022. \\

\appendix

\section{Machine Learning algorithm}
\label{ML}

In this section we want to show  with some detail the characteristics of our Machine Learning algorithms and how they can help us to find stable vacua in a flux string compactification. We make use of two specific algorithms called the Simulated Annealing (SA) and the Conjugate Gradient (CG).
\subsection{Simulated Annealing}
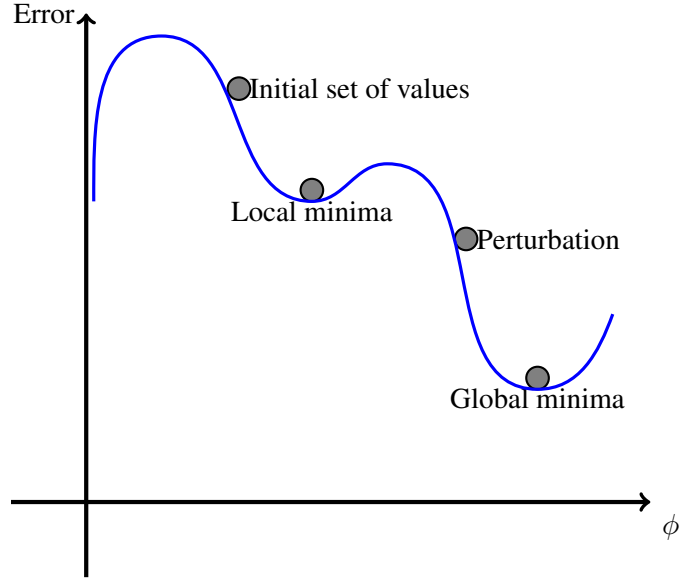
\begin{figure}[htbp]
\center
\begin{tikzpicture}[domain=0:2]

\draw[->,ultra thick] (-1,0) -- (7.5,0) node[below right] {$\phi$};
\draw[->,ultra thick] (0,-1) -- (0,6.5) node[left] {$\text{Error}$};
\draw [fill=gray,thick] (2.03,5.5) circle [radius=0.15] node[right] {Initial set of values};
\draw [fill=gray,thick] (3.00,4.15) circle [radius=0.15] node[below] {Local minima};
\draw [fill=gray,thick] (5.05,3.5) circle [radius=0.15] node[right] {Perturbation};
\draw [fill=gray,thick] (6.00,1.65) circle [radius=0.15] node[below] {Global minima};
\draw[blue,very thick] (0.1,4.0) to [out=90,in=180] (1.0,6.2) to [out=0,in=180] (3.0,4.0) to [out=0,in=180] (4.0,4.5) to [out=0,in=180] (6.0,1.5) to [out=0,in=250] (7.0,2.5);

\end{tikzpicture}
\label{fig:SA}
\caption{In a schematic view, the SA algorithm starts at an arbitrary point in the parameter space of the Error function, then it lets to find in a random manner a best solution leading to a local minima. Once the local minima is reached, the algorithm shall find for alternative paths that finds a better minima by perturbing the solution with a probability $\mathbb{P} \left[ \Delta \text{Error}\right]$ avoiding to get stuck in a local minima.}
\end{figure}

The SA algorithm is one of the most preferred heuristic methods for solving the optimization problems. SA was introduced by inspiring the annealing procedure of the metal working. In general manner, SA algorithm adopts an iterative movement according to a variable parameter which imitates the annealing transaction of the metals. Thus, by taking the objective function as "Error", the SA takes the probability distribution with support $\Delta \text{Error}$ used to replace a new solution as
\begin{equation}
\mathbb{P} \left[ \Delta \text{Error} \right] = \exp \left( \frac{\Delta \text{Error} \log (i+1)}{10} \right) \,,
\end{equation}
for $\Delta \text{Error}$ the change in the error function which depends on an arbitrary number of parameters such as moduli vev's and numerical coefficients that depends on fluxes as well as the non-BPS states, and $i$ the current iteration. Thus as $\Delta \text{Error}$ or the iteration $i$  becomes large the probability to replace a new solution decreases. The SA takes an initial value $\phi_{1}$ and check if 
\begin{equation}
 \text{Error} \left( \phi_1 \right) \leq  \text{Error} \left( \phi_{\text{best}} \right) \,,
\end{equation}
if true, $\phi_{\text{best}}$ is replaced by $\phi_1$, otherwise it is replaced with a probability $\mathbb{P} \left[ \Delta \text{Error} \right]$. A schematic picture of the SA algorithm is shown in Figure \ref{fig:SA}.

\subsection{Conjugate gradient}
Conjugate gradient (CG) is a second-order iterative optimization algorithm designed to find a local minimum provided that the first derivative is known (other alternatives which give us a similar result is the Powells algorithm). The main idea consist in to take repeated steps in conjugate directions of the scalar potential at a given point of the moduli space, since this is the direction of steepest descent. Conversely, stepping in the direction of the gradient which leads to a local minima of the scalar potential. Thus, if the Error function near to the global minima is approximated by
\eq{
\text{Error} (\phi^{i+1} ) = - b_a \phi_a^i + \frac{1}{2} \phi_a^i  \phi_b^{i} A_{ab}
}
the residual defined as
\eq{
r_a \left( \phi \right) =  b_a^{i}- A_{ak} \phi_k^{i} \,,
}
implies that $\partial_ a \text{Error} (\phi^{i+1} ) = - r_a^i$ vanishes at an extremum. Now, in order to move to the minima of the error function, the changes in the gradient have to follow the direction along 
\eq{
 A_{ab} \partial_{a} \text{Error}  (\phi^{i} )  \partial_{b} \text{Error}  (\phi^{i} ) = 0 \,.
 }

This implies that the directions $\partial_a \text{Error}  (\phi^{i} )$ and $\partial_b \text{Error}  (\phi^{i} )$ have to be conjugated. Thus, the CG moves through a conjugate direction leading to a local minima for convex problems. By starting with an initial vector $\phi^0_a$  the conjugate gradient method find two sequences of vectors as
\eq{
\phi^{i+1}_a &= \phi^{i}_a - s \partial_a \text{Error}^i \\
\partial_ a \text{Error}^{i+1} &= - \partial_a  \text{Error}^{i+1} + \gamma  \partial_a  \text{Error}^{i} 
}
where $A_{ab} \partial_a \text{Error}^{i}   \partial_b \text{Error}^{j}  = 0$ for $j < i$, $s$ is an small parameters and
\eq{
\gamma = \frac{g^{i+1}_a g^{i+1}_a}{g^{i}_a g^{i}_a} 
}
is chosen in order to guarantee that the gradients in successive iteration steeps are conjugated.
\subsection{Error functions}
 The objetive function can be written as
 \eq{
 \text{Error} = \sum_{i=1} \alpha_i \cdot \text{error}_i
 }
 for $\alpha_i \in \mathbb{R}$ and in a range of $\left( 0,10^4\right)$\footnote{Notice that this way to implement penalty functions is known as regularization in machine learning and is equivalent to implement Lagrange multipliers in an approximate manner, this is inside the bounds of the convergence criteria.}. This parameters is employed to give a penalty to regions on the moduli space that are not of interest. For instance, if we are looking for dS vacua, the error induced by finding a AdS is weighted by this factor, forcing the algorithm to look for another direction. Thus, in the present work we are interested in finding dS vacua free of negative mass square moduli. Thus our penalty functions shall require 1) to avoid tachyons, 2) to avoid AdS vacua, 3) require that A$_{D5}$ to be positive. In order to penalize this constraints the following errors are employed
 \begin{itemize}
 
\item As we are looking for extrema of the scalar potential, the first error contribution is related to the derivative of the scalar potential. This is applied as
 \eq{
 \text{error}_1 = (\partial_j V )^2
 }
 \item The second contribution of the errors is defined by proposing a continuous function that penalize the error function each time that the parameter space is in a AdS vacua. This is,
\eq{
\text{error}_2 =|V|-V
}
\item The third contribution to the error function is proposed in order to avoid tachyons in the spectrum. For the simple case of two real moduli, the positive mass square moduli require that $\tr m_{ij}^2 > 0$ as well as $\det m ^2 - \frac{1}{4} ( \tr m )^2 > 0$. Thus the third contribution of the error is defined as
\eq{
\text{error}_3 = |\tr m_{ij}| - \tr m_{ij} \,.
}
as well as 
\eq{
\text{error}_4 =  |\det m_{ij}^2 - \frac{1}{4} \tr m_{ij}^2| + \det m_{ij}^2 - \frac{1}{4} \tr m_{ij}^2 \,.
}
\item The fifth contribution to the error is associated to a penalization of the error function each time that the algorithm moves into the region of $A_{D5} <0$. This requirement is implemented in the algorithm as
\eq{
\text{error}_5 = | A_{D5} | -A_{D5} \,.
}
\end{itemize}

\section{More generic vacua}
\begin{figure}[htbp]
   \centering
   \includegraphics[scale=0.5]{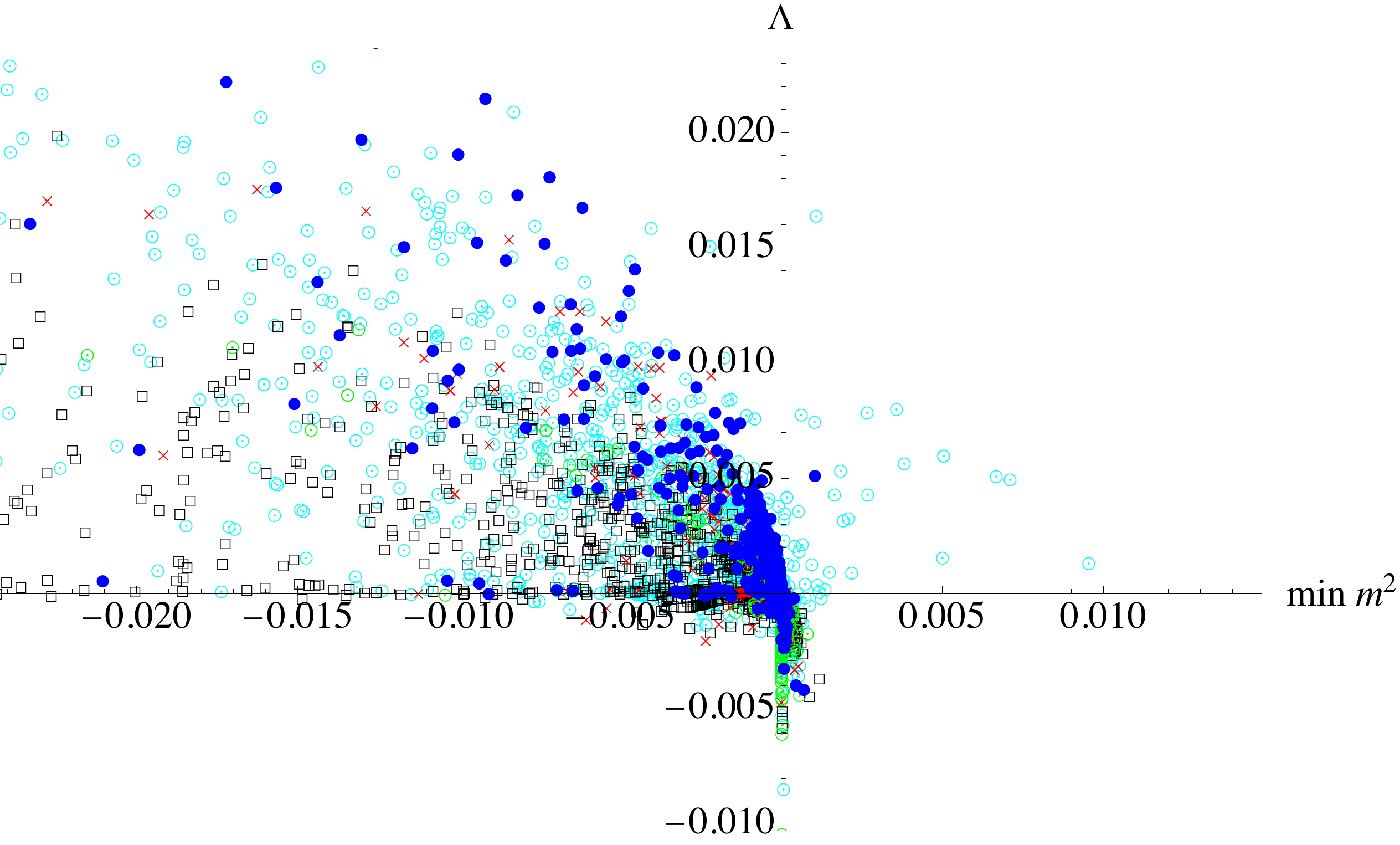} % requires the graphicx package
   \caption{Landscape found by the hybrid method. The red points are found by using a scalar potential with O3, the blue dots are the vacua found by using O7, the green circles are the critical points found by using O5, the cyan circles are found by employing RR F$_5$ fluxes and $\hat D 5$ and the black squares are found by using RR F$_5$ fluxes but not $\hat D 5$.}
   \label{fig:landscape}
\end{figure}

The implementation of our ML algorithms allows to look for stable vacua in more generic conditions. Here we present numerical results by considering extra terms in the scalar potential without wondering wether they can be constructed or not in a consistent scenario. Specifically, we incorporate the contributions to the scalar potential from $O5$ and $O7$ planes and the internal curvature $R_6$ besides the usual 3-form fluxes, the $O3$-plane , $D3$-branes and the non-BPS $\hat{D5}$-branes. Our results are shown in Figure \ref{fig:landscape} where we have plotted each vacua in function of the energy value at the extreme point and the value of the minimal mass eigenvalue. We observe that the majority of the vacua are unstable but some of them are actually dS and stable.\\

All the cases explored in this landscape contain a contribution of the curvature $R_6$ of the internal space. In order to check the landscape of critical points we employ different configurations with a different content of fluxes and O-planes. Some particular comments for each case follows:
\begin{itemize}
\item For the case with $F_3, H_3$, $O3$ and $O5$, the algorithm is able to find stable dS minima. However, almost all the critical points are unstable.
\item For the case of $F_3, H_3$, $O5$ and $O7$, the algorithm was able to find a few stable dS minima. 
\item For the case of $F_3, H_3$, $O3$ and $O5$ the algorithm was not able to find  any dS minima. 
\item For the case of $F_3, H_3, F_5, O3$ and $\hat{D5}$ the algorithm was able to find several dS minima. In particular, it is observed an abundance of dS superior to all the other cases. Besides, as the $\hat D 5$ contribution is removed (black squares), the algorithm was not able to find any dS minima. This suggest that  non-BPS states play an important role in stabilizing the vacua.
\end{itemize}

%Now, we propose that the tadpole is directly related to the term A$_{O3}$. This is since its contribution shall be absent in a compactification with fluxes and localized sources, it must cancel some how in te action. Thus, for instance the CS action
%\eq{
%S_{\text{loc}} = T_{D3} \int_{\mathcal{M}_4} \sqrt {P \left[ g\right]_4} +T_{O3} \int_{\mathcal{M}_4} \sqrt {P \left[ g\right]_4} + \mu_3 (N_{D3} -N_{O3} ) \int_{\mathcal{M}_4} C_4
%}
%whereas, the flux contribution is
%\eq{
%S_{\text{Flux}} = -\frac{\text{i}}{2^3  \tilde \kappa_{10}^2} \int_{\mathcal{M}_4} G_3 \wedge \bar G_3 
%}
%thus, the DBI term effectively contributes to the scalar potential whereas the flux contribution cancels with the flux contribution. 

%Now, in Figure \ref{fig:ao3}, we show the distribution of stable vacua, and its relation with the numerical contribution of A$_{O3}$. As is observed, the red points represent the AdS vacua whereas the blue points are the dS vacua. Thus, the results suggest that the stable dS critical points can only be possible as long as there is a contribution of A$_{O3}$ (like the $\hat D3$ contribution in the KKLT scenario), but the $\hat D 5$ terms are required for stability. \\
%%\begin{figure}[htbp]
%   \centering
%   \includegraphics[scale=0.5]{tadpole.pdf} % requires the graphicx package
%   \caption{Plot of the relation between the stable vacua either AdS or dS and the A$_{O4}$ contribution to the scalar potential coming from the DBI term. At this level, we canno say nothing about the tadpole.}
%   \label{fig:ao3}
%\end{figure}

Finally and just  for sake of comparison, we want to show that the implementation of a penalty constraints in the ML algorithms really impacts on the number of stable vacua we find. Let us look for critical point with the same algorithm and by considering the same content the fluxes as in the body of the paper, i.e., $F_3, H_3$, and $F_5$,  as well as $O3$-planes and non-BPS $\hat{D5}$-branes  (no curvature term). In this case we realize that
\begin{itemize}
\item As we remove the constraints the algorithm finds a lot of critical points but just 6 stable dS against 529 stable AdS. This case is similar to the one obtained by employing the GA+NN classification of our previous work.
\item As we implement the penalty functions, the algorithm is able to find 203 stable dS and 170 stable AdS. This is shown in Figure \ref{fig:ao3}.\\

\begin{figure}[htbp]
   \centering
   \includegraphics[scale=0.5]{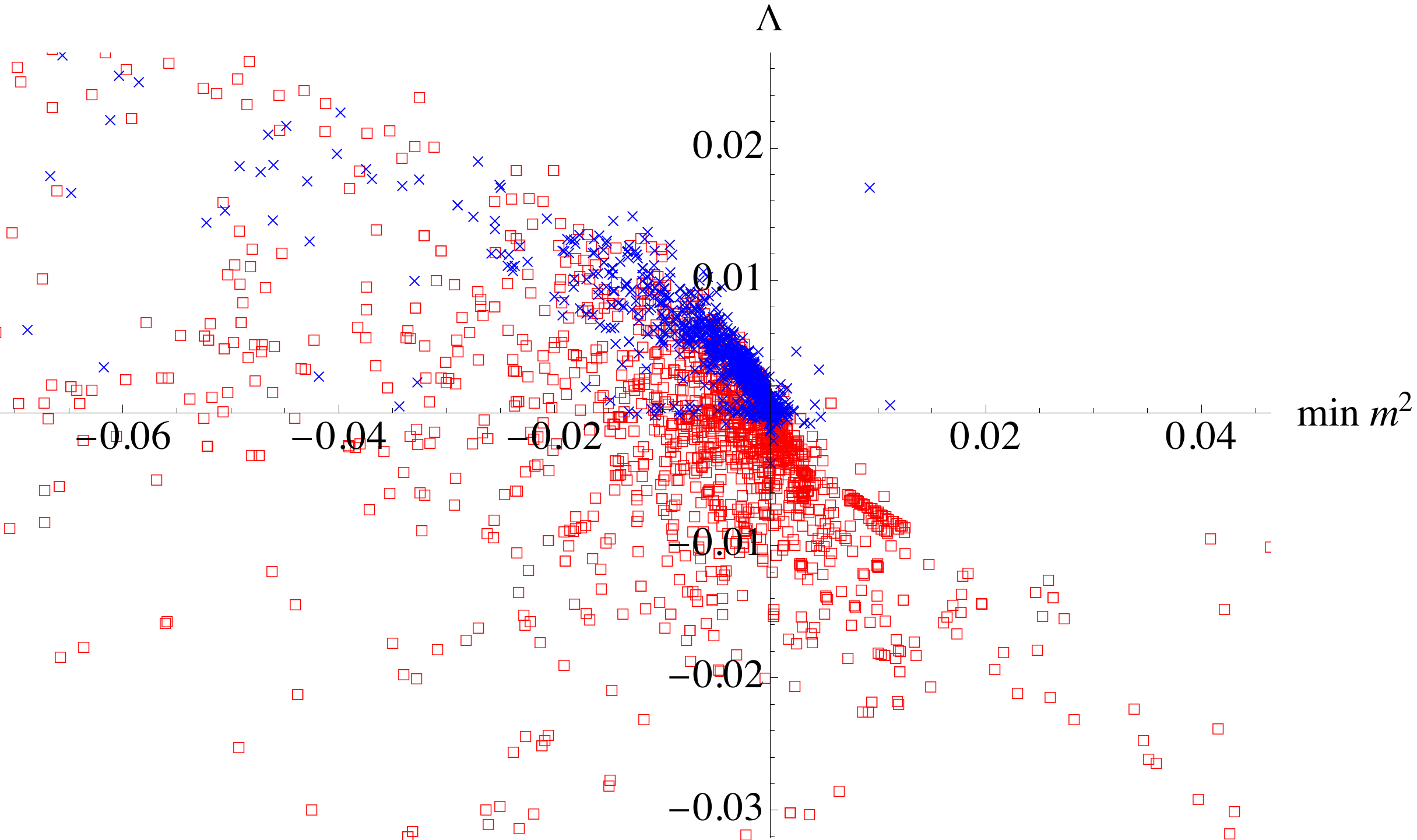} % requires the graphicx package
   \caption{Plot obtained by using non-penalty constraints (red points) and penalty constraints (blue points).}
   \label{fig:ao3}
\end{figure}

\end{itemize}

\bibliographystyle{JHEP}
\bibliography{references2}

\providecommand{\href}[2]{#2}\begingroup\raggedright\begin{thebibliography}{10}

\bibitem{Palti:2019pca}
E.~Palti, \emph{{The Swampland: Introduction and Review}},
  \href{https://doi.org/10.1002/prop.201900037}{\emph{Fortsch. Phys.}
  {\bfseries 67} (2019) 1900037}
  [\href{https://arxiv.org/abs/1903.06239}{{\ttfamily 1903.06239}}].

\bibitem{vanBeest:2021lhn}
M.~van Beest, J.~Calder\'on-Infante, D.~Mirfendereski and I.~Valenzuela,
  \emph{{Lectures on the Swampland Program in String Compactifications}},
  \href{https://arxiv.org/abs/2102.01111}{{\ttfamily 2102.01111}}.

\bibitem{Grana:2021zvf}
M.~Gra\~na and A.~Herr\'aez, \emph{{The Swampland Conjectures: A Bridge from
  Quantum Gravity to Particle Physics}},
  \href{https://doi.org/10.3390/universe7080273}{\emph{Universe} {\bfseries 7}
  (2021) 273} [\href{https://arxiv.org/abs/2107.00087}{{\ttfamily
  2107.00087}}].

\bibitem{Ooguri:2006in}
H.~Ooguri and C.~Vafa, \emph{{On the Geometry of the String Landscape and the
  Swampland}},
  \href{https://doi.org/10.1016/j.nuclphysb.2006.10.033}{\emph{Nucl. Phys. B}
  {\bfseries 766} (2007) 21}
  [\href{https://arxiv.org/abs/hep-th/0605264}{{\ttfamily hep-th/0605264}}].

\bibitem{Garg:2018reu}
S.~K. Garg and C.~Krishnan, \emph{{Bounds on Slow Roll and the de Sitter
  Swampland}}, \href{https://doi.org/10.1007/JHEP11(2019)075}{\emph{JHEP}
  {\bfseries 11} (2019) 075}
  [\href{https://arxiv.org/abs/1807.05193}{{\ttfamily 1807.05193}}].

\bibitem{Ooguri:2018wrx}
H.~Ooguri, E.~Palti, G.~Shiu and C.~Vafa, \emph{{Distance and de Sitter
  Conjectures on the Swampland}},
  \href{https://doi.org/10.1016/j.physletb.2018.11.018}{\emph{Phys. Lett.}
  {\bfseries B788} (2019) 180}
  [\href{https://arxiv.org/abs/1810.05506}{{\ttfamily 1810.05506}}].

\bibitem{Lust:2019zwm}
D.~Lüst, E.~Palti and C.~Vafa, \emph{{AdS and the Swampland}},
  \href{https://doi.org/10.1016/j.physletb.2019.134867}{\emph{Phys. Lett. B}
  {\bfseries 797} (2019) 134867}
  [\href{https://arxiv.org/abs/1906.05225}{{\ttfamily 1906.05225}}].

\bibitem{Apers:2022zjx}
F.~Apers, M.~Montero, T.~Van~Riet and T.~Wrase, \emph{{Comments on classical
  AdS flux vacua with scale separation}},
  \href{https://arxiv.org/abs/2202.00682}{{\ttfamily 2202.00682}}.

\bibitem{Bena:2018fqc}
I.~Bena, E.~Dudas, M.~Gra\~na and S.~L\"ust, \emph{{Uplifting Runaways}},
  \href{https://doi.org/10.1002/prop.201800100}{\emph{Fortsch. Phys.}
  {\bfseries 67} (2019) 1800100}
  [\href{https://arxiv.org/abs/1809.06861}{{\ttfamily 1809.06861}}].

\bibitem{Blumenhagen:2019kqm}
R.~Blumenhagen, M.~Brinkmann and A.~Makridou, \emph{{A Note on the dS Swampland
  Conjecture, Non-BPS Branes and K-Theory}},
  \href{https://doi.org/10.1002/prop.201900068}{\emph{Fortsch. Phys.}
  {\bfseries 67} (2019) 1900068}
  [\href{https://arxiv.org/abs/1906.06078}{{\ttfamily 1906.06078}}].

\bibitem{Damian:2019bkb}
C.~Damian and O.~Loaiza-Brito, \emph{{Some remarks on the dS conjecture, fluxes
  and K-theory in IIB toroidal compactifications}},
  \href{https://arxiv.org/abs/1906.08766}{{\ttfamily 1906.08766}}.

\bibitem{Uranga:2000xp}
A.~M. Uranga, \emph{{D-brane probes, RR tadpole cancellation and K theory
  charge}}, \href{https://doi.org/10.1016/S0550-3213(00)00787-2}{\emph{Nucl.
  Phys. B} {\bfseries 598} (2001) 225}
  [\href{https://arxiv.org/abs/hep-th/0011048}{{\ttfamily hep-th/0011048}}].

\bibitem{Blumenhagen:2021nmi}
R.~Blumenhagen and N.~Cribiori, \emph{{Open-Closed Correspondence of K-theory
  and Cobordism}},  \href{https://arxiv.org/abs/2112.07678}{{\ttfamily
  2112.07678}}.

\bibitem{Hertzberg:2007wc}
M.~P. Hertzberg, S.~Kachru, W.~Taylor and M.~Tegmark, \emph{{Inflationary
  Constraints on Type IIA String Theory}},
  \href{https://doi.org/10.1088/1126-6708/2007/12/095}{\emph{JHEP} {\bfseries
  12} (2007) 095} [\href{https://arxiv.org/abs/0711.2512}{{\ttfamily
  0711.2512}}].

\bibitem{Shiu:2011zt}
G.~Shiu and Y.~Sumitomo, \emph{{Stability Constraints on Classical de Sitter
  Vacua}}, \href{https://doi.org/10.1007/JHEP09(2011)052}{\emph{JHEP}
  {\bfseries 09} (2011) 052} [\href{https://arxiv.org/abs/1107.2925}{{\ttfamily
  1107.2925}}].

\bibitem{He:2018jtw}
Y.-H. He, \emph{{The Calabi-Yau Landscape: from Geometry, to Physics, to
  Machine-Learning}},  \href{https://arxiv.org/abs/1812.02893}{{\ttfamily
  1812.02893}}.

\bibitem{Ashmore:2019wzb}
A.~Ashmore, Y.-H. He and B.~A. Ovrut, \emph{{Machine learning Calabi-Yau
  metrics}},  \href{https://arxiv.org/abs/1910.08605}{{\ttfamily 1910.08605}}.

\bibitem{Parr:2019bta}
E.~Parr and P.~K.~S. Vaudrevange, \emph{{Contrast data mining for the MSSM from
  strings}}, \href{https://doi.org/10.1016/j.nuclphysb.2020.114922}{\emph{Nucl.
  Phys.} {\bfseries B952} (2020) 114922}
  [\href{https://arxiv.org/abs/1910.13473}{{\ttfamily 1910.13473}}].

\bibitem{Bao:2020sqg}
J.~Bao, Y.-H. He, E.~Hirst and S.~Pietromonaco, \emph{{Lectures on the
  Calabi-Yau Landscape}},  \href{https://arxiv.org/abs/2001.01212}{{\ttfamily
  2001.01212}}.

\bibitem{Halverson:2020opj}
J.~Halverson and C.~Long, \emph{{Statistical Predictions in String Theory and
  Deep Generative Models}},  \href{https://arxiv.org/abs/2001.00555}{{\ttfamily
  2001.00555}}.

\bibitem{Gal:2020dyc}
Y.~Gal, V.~Jejjala, D.~K. Mayorga~Pena and C.~Mishra, \emph{{Baryons from
  Mesons: A Machine Learning Perspective}},
  \href{https://arxiv.org/abs/2003.10445}{{\ttfamily 2003.10445}}.

\bibitem{Erbin:2020tks}
H.~Erbin and R.~Finotello, \emph{{Machine learning for complete intersection
  Calabi-Yau manifolds: a methodological study}},
  \href{https://doi.org/10.1103/PhysRevD.103.126014}{\emph{Phys. Rev. D}
  {\bfseries 103} (2021) 126014}
  [\href{https://arxiv.org/abs/2007.15706}{{\ttfamily 2007.15706}}].

\bibitem{CaboBizet:2020cse}
N.~Cabo~Bizet, C.~Damian, O.~Loaiza-Brito, D.~K.~M. Pe\~na and J.~A. Monta\~nez
  Barrera, \emph{{Testing Swampland Conjectures with Machine Learning}},
  \href{https://doi.org/10.1140/epjc/s10052-020-8332-9}{\emph{Eur. Phys. J. C}
  {\bfseries 80} (2020) 766}
  [\href{https://arxiv.org/abs/2006.07290}{{\ttfamily 2006.07290}}].

\bibitem{Cole:2021nnt}
A.~Cole, S.~Krippendorf, A.~Schachner and G.~Shiu, \emph{{Probing the Structure
  of String Theory Vacua with Genetic Algorithms and Reinforcement Learning}},
  in \emph{{35th Conference on Neural Information Processing Systems}}, 11,
  2021, \href{https://arxiv.org/abs/2111.11466}{{\ttfamily 2111.11466}}.

\bibitem{He:2022cpz}
Y.-H. He, \emph{{From the String Landscape to the Mathematical Landscape: a
  Machine-Learning Outlook}},  in \emph{{14th International Workshop on Lie
  Theory and Its Applications in Physics}}, 2, 2022,
  \href{https://arxiv.org/abs/2202.06086}{{\ttfamily 2202.06086}}.

\bibitem{Plauschinn:2018wbo}
E.~Plauschinn, \emph{{Non-geometric backgrounds in string theory}},
  \href{https://doi.org/10.1016/j.physrep.2018.12.002}{\emph{Phys. Rept.}
  {\bfseries 798} (2019) 1} [\href{https://arxiv.org/abs/1811.11203}{{\ttfamily
  1811.11203}}].

\bibitem{Danielsson:2012et}
U.~H. Danielsson, G.~Shiu, T.~Van~Riet and T.~Wrase, \emph{{A note on obstinate
  tachyons in classical dS solutions}},
  \href{https://doi.org/10.1007/JHEP03(2013)138}{\emph{JHEP} {\bfseries 03}
  (2013) 138} [\href{https://arxiv.org/abs/1212.5178}{{\ttfamily 1212.5178}}].

\bibitem{Blumenhagen:2015xpa}
R.~Blumenhagen, C.~Damian, A.~Font, D.~Herschmann and R.~Sun, \emph{{The
  Flux-Scaling Scenario: De Sitter Uplift and Axion Inflation}},
  \href{https://doi.org/10.1002/prop.201600030}{\emph{Fortsch. Phys.}
  {\bfseries 64} (2016) 536}
  [\href{https://arxiv.org/abs/1510.01522}{{\ttfamily 1510.01522}}].

\bibitem{Junghans:2016uvg}
D.~Junghans, \emph{{Tachyons in Classical de Sitter Vacua}},
  \href{https://doi.org/10.1007/JHEP06(2016)132}{\emph{JHEP} {\bfseries 06}
  (2016) 132} [\href{https://arxiv.org/abs/1603.08939}{{\ttfamily
  1603.08939}}].

\bibitem{CaboBizet:2016qsa}
N.~Cabo~Bizet and S.~Hirano, \emph{{Revisiting constraints on uplifts to de
  Sitter vacua}},  \href{https://arxiv.org/abs/1607.01139}{{\ttfamily
  1607.01139}}.

\bibitem{Andriot:2018ept}
D.~Andriot, \emph{{New constraints on classical de Sitter: flirting with the
  swampland}}, \href{https://doi.org/10.1002/prop.201800103}{\emph{Fortsch.
  Phys.} {\bfseries 67} (2019) 1800103}
  [\href{https://arxiv.org/abs/1807.09698}{{\ttfamily 1807.09698}}].

\bibitem{Kallosh:2018nrk}
R.~Kallosh and T.~Wrase, \emph{{dS Supergravity from 10d}},
  \href{https://doi.org/10.1002/prop.201800071}{\emph{Fortsch. Phys.}
  {\bfseries 67} (2019) 1800071}
  [\href{https://arxiv.org/abs/1808.09427}{{\ttfamily 1808.09427}}].

\bibitem{Andriot:2018mav}
D.~Andriot and C.~Roupec, \emph{{Further refining the de Sitter swampland
  conjecture}}, \href{https://doi.org/10.1002/prop.201800105}{\emph{Fortsch.
  Phys.} {\bfseries 67} (2019) 1800105}
  [\href{https://arxiv.org/abs/1811.08889}{{\ttfamily 1811.08889}}].

\bibitem{Andriot:2019wrs}
D.~Andriot, \emph{{Open problems on classical de Sitter solutions}},
  \href{https://doi.org/10.1002/prop.201900026}{\emph{Fortsch. Phys.}
  {\bfseries 67} (2019) 1900026}
  [\href{https://arxiv.org/abs/1902.10093}{{\ttfamily 1902.10093}}].

\bibitem{Andriot:2020wpp}
D.~Andriot, P.~Marconnet and T.~Wrase, \emph{{New de Sitter solutions of 10d
  type IIB supergravity}},
  \href{https://doi.org/10.1007/JHEP08(2020)076}{\emph{JHEP} {\bfseries 08}
  (2020) 076} [\href{https://arxiv.org/abs/2005.12930}{{\ttfamily
  2005.12930}}].

\bibitem{Andriot:2021rdy}
D.~Andriot, \emph{{Tachyonic de Sitter Solutions of 10d Type II
  Supergravities}},
  \href{https://doi.org/10.1002/prop.202100063}{\emph{Fortsch. Phys.}
  {\bfseries 69} (2021) 2100063}
  [\href{https://arxiv.org/abs/2101.06251}{{\ttfamily 2101.06251}}].

\bibitem{McAllister:2008hb}
L.~McAllister, E.~Silverstein and A.~Westphal, \emph{{Gravity Waves and Linear
  Inflation from Axion Monodromy}},
  \href{https://doi.org/10.1103/PhysRevD.82.046003}{\emph{Phys. Rev. D}
  {\bfseries 82} (2010) 046003}
  [\href{https://arxiv.org/abs/0808.0706}{{\ttfamily 0808.0706}}].

\bibitem{Cai:2014vua}
Y.-F. Cai, F.~Chen, E.~G.~M. Ferreira and J.~Quintin, \emph{{New model of axion
  monodromy inflation and its cosmological implications}},
  \href{https://doi.org/10.1088/1475-7516/2016/06/027}{\emph{JCAP} {\bfseries
  06} (2016) 027} [\href{https://arxiv.org/abs/1412.4298}{{\ttfamily
  1412.4298}}].

\bibitem{Grimm:2004uq}
T.~W. Grimm and J.~Louis, \emph{{The Effective action of N = 1 Calabi-Yau
  orientifolds}},
  \href{https://doi.org/10.1016/j.nuclphysb.2004.08.005}{\emph{Nucl. Phys. B}
  {\bfseries 699} (2004) 387}
  [\href{https://arxiv.org/abs/hep-th/0403067}{{\ttfamily hep-th/0403067}}].

\bibitem{Witten:1998cd}
E.~Witten, \emph{{D-branes and K theory}},
  \href{https://doi.org/10.1088/1126-6708/1998/12/019}{\emph{JHEP} {\bfseries
  12} (1998) 019} [\href{https://arxiv.org/abs/hep-th/9810188}{{\ttfamily
  hep-th/9810188}}].

\bibitem{Asakawa:2002nv}
T.~Asakawa, S.~Sugimoto and S.~Terashima, \emph{{D branes and KK theory in type
  1 string theory}},
  \href{https://doi.org/10.1088/1126-6708/2002/05/007}{\emph{JHEP} {\bfseries
  05} (2002) 007} [\href{https://arxiv.org/abs/hep-th/0202165}{{\ttfamily
  hep-th/0202165}}].

\bibitem{Garcia-Compean:2008fzo}
H.~Garcia-Compean, W.~Herrera-Suarez, B.~A. Itza-Ortiz and O.~Loaiza-Brito,
  \emph{{D-Branes in Orientifolds and Orbifolds and Kasparov KK-Theory}},
  \href{https://doi.org/10.1088/1126-6708/2008/12/007}{\emph{JHEP} {\bfseries
  12} (2008) 007} [\href{https://arxiv.org/abs/0809.4238}{{\ttfamily
  0809.4238}}].

\bibitem{Loaiza-Brito:2001yer}
O.~Loaiza-Brito and A.~M. Uranga, \emph{{The Fate of the type I nonBPS
  D7-brane}}, \href{https://doi.org/10.1016/S0550-3213(01)00505-3}{\emph{Nucl.
  Phys. B} {\bfseries 619} (2001) 211}
  [\href{https://arxiv.org/abs/hep-th/0104173}{{\ttfamily hep-th/0104173}}].

\bibitem{Bergman:2001rp}
O.~Bergman, E.~G. Gimon and S.~Sugimoto, \emph{{Orientifolds, RR torsion, and K
  theory}}, \href{https://doi.org/10.1088/1126-6708/2001/05/047}{\emph{JHEP}
  {\bfseries 05} (2001) 047}
  [\href{https://arxiv.org/abs/hep-th/0103183}{{\ttfamily hep-th/0103183}}].

\bibitem{Freivogel:2016qwc}
B.~Freivogel and M.~Kleban, \emph{{Vacua Morghulis}},
  \href{https://arxiv.org/abs/1610.04564}{{\ttfamily 1610.04564}}.

\bibitem{Ooguri:2016pdq}
H.~Ooguri and C.~Vafa, \emph{{Non-supersymmetric AdS and the Swampland}},
  \href{https://doi.org/10.4310/ATMP.2017.v21.n7.a8}{\emph{Adv. Theor. Math.
  Phys.} {\bfseries 21} (2017) 1787}
  [\href{https://arxiv.org/abs/1610.01533}{{\ttfamily 1610.01533}}].

\bibitem{Frau:1999qs}
M.~Frau, L.~Gallot, A.~Lerda and P.~Strigazzi, \emph{{Stable nonBPS D-branes in
  type I string theory}},
  \href{https://doi.org/10.1016/S0550-3213(99)00624-0}{\emph{Nucl. Phys. B}
  {\bfseries 564} (2000) 60}
  [\href{https://arxiv.org/abs/hep-th/9903123}{{\ttfamily hep-th/9903123}}].

\bibitem{Lerda:1999um}
A.~Lerda and R.~Russo, \emph{{Stable nonBPS states in string theory: A
  Pedagogical review}},
  \href{https://doi.org/10.1142/S0217751X00000380}{\emph{Int. J. Mod. Phys. A}
  {\bfseries 15} (2000) 771}
  [\href{https://arxiv.org/abs/hep-th/9905006}{{\ttfamily hep-th/9905006}}].

\bibitem{Brown:2010mf}
A.~R. Brown and A.~Dahlen, \emph{{Bubbles of Nothing and the Fastest Decay in
  the Landscape}},
  \href{https://doi.org/10.1103/PhysRevD.84.043518}{\emph{Phys. Rev. D}
  {\bfseries 84} (2011) 043518}
  [\href{https://arxiv.org/abs/1010.5240}{{\ttfamily 1010.5240}}].

\bibitem{Candelas:2014jma}
P.~Candelas, A.~Constantin, C.~Damian, M.~Larfors and J.~F. Morales,
  \emph{{Type IIB flux vacua from G-theory I}},
  \href{https://doi.org/10.1007/JHEP02(2015)187}{\emph{JHEP} {\bfseries 02}
  (2015) 187} [\href{https://arxiv.org/abs/1411.4785}{{\ttfamily 1411.4785}}].

\bibitem{Candelas:2014kma}
P.~Candelas, A.~Constantin, C.~Damian, M.~Larfors and J.~F. Morales,
  \emph{{Type IIB flux vacua from G-theory II}},
  \href{https://doi.org/10.1007/JHEP02(2015)188}{\emph{JHEP} {\bfseries 02}
  (2015) 188} [\href{https://arxiv.org/abs/1411.4786}{{\ttfamily 1411.4786}}].

\end{thebibliography}\endgroup

\end{document}